\documentclass{ws-mpla}

\begin{document}

\def\d{{\rm d}}
\def\p{I\!\!P}

\def\lp{\left. }
\def\rp{\right. }
\def\lr{\left( }
\def\rr{\right) }
\def\le{\left[ }
\def\re{\right] }
\def\lg{\left\{ }
\def\rg{\right\} }
\def\lb{\left| }
\def\rb{\right| }

\def\beq{\begin{equation}}
\def\eeq{\end{equation}}
\def\bea{\begin{eqnarray}}
\def\eea{\end{eqnarray}}

\markboth{M.\ Klasen, G.\ Kramer}
{Review of factorization breaking in diffractive photoproduction of dijets}

\catchline{}{}{}{}{}

\title{Review of factorization breaking in diffractive photoproduction of 
       dijets}

\author{\footnotesize MICHAEL KLASEN}

\address{Laboratoire de Physique Subatomique et de Cosmologie,
 Universit\'e Joseph Fourier / CNRS-IN2P3 / INPG, 53 Avenue des Martyrs,
 F-38026 Grenoble, France\\
 klasen@lpsc.in2p3.fr}

\author{GUSTAV KRAMER}

\address{II.\ Institut f\"ur Theoretische Physik, Universit\"at
 Hamburg, Luruper Chaussee 149, \\ D-22761 Hamburg, Germany
}

\maketitle

\pub{Received (Day Month Year)}{Revised (Day Month Year)}

\begin{abstract}
 After the final analyses of the H1 and ZEUS collaborations for the diffractive
 photoproduction of dijets have appeared, we have recalculated these cross
 sections in next-to-leading order (NLO) of perturbative QCD to see whether they
 can be interpreted consistently. The results of these calculations are compared
 to the data of both collaborations. We find that at NLO the cross sections
 disagree with the data, showing that factorization breaking occurs at this order.
 If direct and resolved contributions are both suppressed by the same amount, the
 global suppression factor depends on the transverse-energy cut and is $0.42$ for
 the H1 and $0.71$ for the ZEUS analysis. However, by suppressing only the
 resolved contribution by a factor of approximately three, also reasonably good
 agreement with all the data is found. The size of the factorization breaking
 effects for resolved photons agrees with absorptive-model predictions.

\end{abstract}

\ccode{12.38.Bx; 12.39.St; 12.40.Nn; 13.87.Ce.}

\vspace*{-160mm}
\noindent DESY 08-074\\
\noindent LPSC 08-070
\vspace*{150mm}

\section{Introduction}

\begin{figure}
 \centering
 \includegraphics[width=0.6\columnwidth]{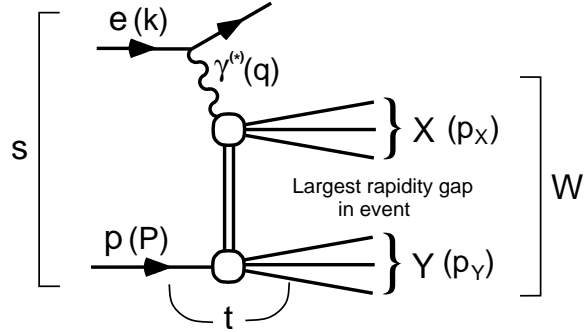}
 \caption{\label{fig:0}Diffractive scattering process $ep\to eXY$, where
 the hadronic systems $X$ and $Y$ are separated by the largest rapidity
 gap in the final state.}
\end{figure}

It is well known that at high-energy colliders such as the $ep$ collider HERA
at DESY and the $p\bar{p}$ collider Tevatron at Fermilab, a large fraction of
the observed events are diffractive. These events are defined experimentally
by the presence of a forward-going hadronic system $Y$ with four-momentum $p_Y$,
low mass $M_Y$ (typically a proton that remained intact or a proton plus
low-lying nucleon resonances), small four-momentum transfer $t=(P-p_Y)^2$, and
small longitudinal-momentum transfer $x_{\p} = q(P-p_Y)/(qP)$ from the incoming
proton with four momentum $P$ to the central hadronic system $X$ (see Fig.\
\ref{fig:0} for the case of $ep \rightarrow eXY$). Experimentally, a large
%
%
rapidity gap separates the hadronic system $X$ with invariant mass $M_X$ from the
final-state system $Y$ with invariant mass $M_Y$.

Theoretically, diffractive interactions are described in the framework of Regge
theory \cite{1} as the exchange of a trajectory with vacuum quantum numbers,
the pomeron ($\p$) trajectory. Then the object exchanged between the systems
$X$ and $Y$, as indicated in Fig.\ \ref{fig:0}, is the pomeron (or additional
lower-lying Regge poles), and the upper vertex of the process $e\p \rightarrow
eX$ can be interpreted as deep-inelastic scattering (DIS) on the pomeron target
for the case that the virtuality of the exchanged photon $Q^2 =-q^2$ is
sufficiently large. In analogy to DIS on a proton target, $ep \rightarrow eX$,
the cross section for the process $e\p \rightarrow eX$ in the DIS region can be
expressed as the convolution of partonic cross sections and universal parton
distribution functions (PDFs) of the pomeron. The partonic cross sections are
the same as for DIS $ep$ scattering. Usually these pomeron PDFs are multiplied
with vertex functions for the lower vertex in Fig.\ \ref{fig:0}, yielding the
diffractive parton distribution functions (DPDFs). The $Q^2$-evolution of the
DPDFs is calculated with the usual DGLAP \cite{2} evolution equations known
from $ep \rightarrow eX$ DIS. Except for their evolution with $Q^2$, the DPDFs
can not be calculated in the framework of perturbative QCD and must be
determined from experiment. Such DPDFs \cite{3,4,5,6} have been obtained from
the HERA inclusive measurements of the diffractive structure function $F^D_2$
\cite{3,4}, defined analogously to the proton structure function $F_2$.

Similarly to diffractive DIS, $ep \rightarrow eXY$, where the presence of the
large scale $Q$ allows for the application of perturbative QCD and $X$ comprises
the sum over all possible final states, many other processes with a hard scale
provided by specific final states in the central system $X$ can be predicted
using QCD perturbation theory. Such processes, usually called hard diffractive
processes, are e.g.\ dijet production in diffractive photoproduction ($Q^2
\simeq 0$) and DIS ($Q^2 \neq 0$), where the large scale is given by the jet
transverse energy $E_T^{jet}$ and possibly $Q$, and diffractive open
heavy-flavor production, where the large scale is given by the heavy-flavor mass
and possibly $E_T$ and/or $Q$, in photoproduction or DIS and many more
diffractive processes induced by $p\bar p$ or $pp$ collisions. The central
problem in hard diffraction is the problem of QCD factorization, i.e.\ the
question whether diffractive cross sections are factorisable into universal
DPDFs and partonic cross sections, which are calculable in perturbative QCD.
This question is the subject of the current 
debate in diffractive physics and is of particular interest for the prospects
of discovery of new particles such as the Higgs boson in diffractive reactions 
at the LHC \cite{7,8,9}.

For the inclusive DIS process, factorization has indeed been proven to hold
\cite{10}, and on this basis DPDFs have been extracted at $Q^2 \neq 0$ 
\cite{3,4,5} from high-precision inclusive measurements of the process
$ep \rightarrow eXY$ using the usual DGLAP evolution equations. The proof of 
the factorization formula, usually referred to as the validity of QCD 
factorization in hard diffraction, also appears to be valid for the production
of specific final states in DIS, as e.g.\ the production of jets or
heavy-flavor particles, and for the direct part of photoproduction 
($Q^2 \simeq 0$) or low-$Q^2$ electroproduction of jets \cite{10}. However,
factorization does not hold for hard processes in diffractive hadron-hadron
scattering. The problem is that soft interactions between the ingoing hadrons
and/or their remnants occur in both the initial and the final state. This 
agrees with experimental measurements at the Tevatron \cite{11}. Predictions
of diffractive dijet cross sections for collisions as measured by CDF using 
DPDFs determined earlier by the H1 collaboration \cite{12} at HERA overestimate
the measured cross section by up to an order of magnitude \cite{11}. This
large suppression of the CDF cross section can be explained by the rescattering
of the two incoming hadron beams, which, by creating additional hadrons, destroy
the rapidity gap \cite{13}.

\begin{figure}
 \centering
 \includegraphics[width=0.46\columnwidth]{fig1a}
 \includegraphics[width=0.49\columnwidth]{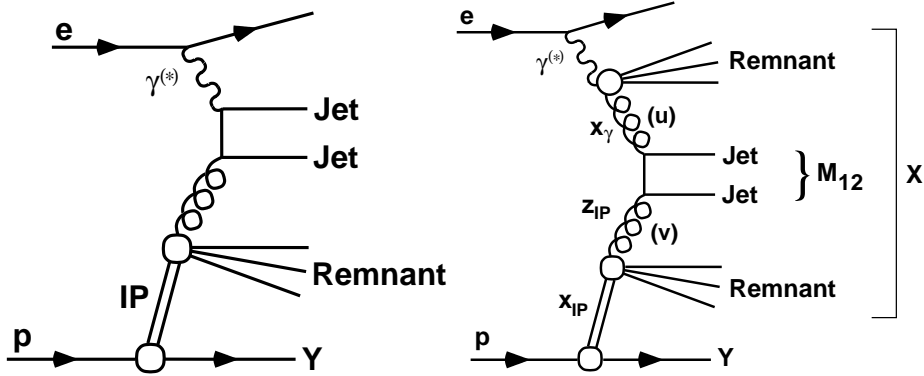}
 \caption{\label{fig:1}Diffractive production of dijets with invariant mass
 $M_{12}$ in direct (left) and resolved (right) photon-pomeron collisions,
 leading to the production of one or two additional remnant jets.}
\end{figure}

Jet production with real photons involves direct interactions of the photon
with quarks or gluons originating from the proton or pomeron, respectively,
as well as resolved photon contributions leading to parton-parton interactions
with an additional remnant jet coming from the photon as reviewed in \cite{14}
(see Fig.\ \ref{fig:1}). For the direct interactions, we expect factorization
%
%
to be valid as in the case of inclusive DIS, as already mentioned, whereas we
expect it to fail for the resolved process as in hadron-hadron scattering. For
this part of photoproduction we would therefore expect a similar suppression
factor (sometimes also called rapidity-gap survival probability) due to
rescattering effects of the ingoing partons or hadrons. Introducing
vector-meson dominance photon fluctuations, such a suppression by about a
factor of three was predicted for resolved photoproduction at HERA \cite{15}.

The first measurements of dijet cross sections in diffractive photoproduction
have been presented by the H1 collaboration as contributions to two conferences
\cite{16}. The kinematic range for these data were $Q^2 < 0.01$ GeV$^2$, $x_{\p}<
0.03$, $E_T^{jet1} > 5$ GeV, $E_T^{jet2} > 4$ GeV and $165<W<240$ GeV, where jets
were identified using the inclusive $k_T$-cluster algorithm (the definitions of
these and the following variables will be given in the next section). The
measured cross sections as a function of $x_{\gamma}^{obs}$ and $z_{\p}^{obs}$
were compared to leading-order (LO) QCD predictions, using the RAPGAP Monte Carlo
model  \cite{17}. For the DPDFs the LO `H1 2002 fit' was used \cite{12}. The
two cross sections were found to be well described by the predictions in
normalization and shape over the whole range of $x_{\gamma}^{obs}$ and
$z_{\p}^{obs}$, showing no breakdown of factorization neither in resolved nor
in direct photoproduction. In addition, normalized cross sections as a function
of various other variables were compared to the predictions with the result
that all measured distributions were in good agreement.

Subsequently we calculated the next-to-leading order (NLO) corrections for
the cross section of diffractive dijet production using the same
kinematic cuts and with the same DPDFs as in the first H1 analysis
\cite{16} on the basis of our previous work on NLO corrections for
inclusive direct \cite{18} and resolved \cite{19} dijet photoproduction.
While at LO good agreement with the H1 data \cite{16} was found, consistent
with the finding in the H1 analysis \cite{16}, it was found that the NLO
corrections increase the cross section significantly \cite{20,21} and
require a suppression factor of the order of $R=0.5$. Since on theoretical
grounds only a suppression of the resolved cross section would be
acceptable, we demonstrated in \cite{20,21} that by multiplying the
resolved cross section with the suppression factor $R = 0.34$, reasonably
good agreement with the preliminary H1 data \cite{16} could be achieved.
This value for the suppression factor turned out to be in good agreement
with the prediction of \cite{15}.

The first experimental data from the ZEUS collaboration were presented at
the DIS workshop in 2004 \cite{22}. The dijet cross sections were obtained
in the kinematic range $Q^2 < 1$ GeV$^2$, $x_{\p} < 0.035$ and $E_T^{jet1(2)}
> 7.5$ $(6.5)$ GeV. For these kinematic constraints NLO calculations were not
available in 2004. So, the measurements were compared to LO calculations, 
unfortunately with previous H1 DPDFs \cite{23} with the result, that good 
agreement in the shape was achieved. But the normalization was off by a factor of
$0.6$, which was attributed to the older DPDF input \cite{24}, so that the 
H1 and ZEUS results were consistent with each other. The situation concerning 
the agreement of H1 and ZEUS data and the influence of NLO corrections 
improved already considerably in the fall of 2004. These preliminary data from
both HERA collaborations together with comparisons to NLO calculations based on
the DPDF fits from \cite{12} were presented at workshops and conferences in the
following years.

In 2006 the H1 collaboration published their final DPDF fits from their 
high-precision measurements using the DGLAP evolution equations \cite{4}.  This
analysis was based on the larger data sample of the years 1997-2000 as compared
to the earlier preliminary DPDF sets \cite{12}. In \cite{4} two DPDF sets, the
'H1 2006 fit A' and the 'H1 2006 fit B' were presented, which both give a good
description of the inclusive diffractive data. These two sets differ mainly in
the gluon density at large fractional parton momenta, which is poorly 
constrained by the inclusive diffractive scattering data, since there is no
direct coupling of the photon to gluons, so that the gluon density is 
constrained only through the evolution. The gluon density of fit A is peaked at
the starting scale at large fractional momenta, whereas fit B is flat in this
region.

In 2007 the final publications for diffractive dijet production appeared 
\cite{25}. The comparison between these experimental results and the NLO
theory was based on the new and final DPDFs from H1 \cite{4}. The
differential cross sections as measured by H1 \cite{25} were compared to NLO
predictions obtained with the Frixione program \cite{26} interfaced to  the
`H1 2006 fit B' DPDFs. The conclusions deduced earlier from the
comparison with the preliminary data and the preliminary `H1 2002 fit'
\cite{12} are fully confirmed in
\cite{25} with the new DPDFs fits \cite{4}.
In particular, a global suppression is obtained, independent of the
DPDFs fits used, i.e.\ fit A or fit B, by considering the ratio of measured
dijet cross sections to NLO predictions in photoproduction in relation to
the same ratio in DIS. In this comparison the value of the suppression is 
$0.5$. In addition, by using this overall suppression factor, H1 
obtained a good description of all the measured distributions in the variables
$z_{\p}^{obs}$, $x_{\gamma}^{obs}$, $x_{\p}$, $W$, $E_T^{jet1}$,
$\bar{\eta}^{jets}$, $|\Delta \eta^{jets}|$ and $M_{12}$
interfaced with the
`H1 2006 fit B' DPDFs and 
taking into account hadronization corrections \cite{25}. Finally, the 
H1 collaboration investigated how well the data are describable under the 
assumption that in the NLO calculation the cross section for 
$x_{\gamma}^{obs} > 0.9$ is not suppressed. The best agreement
in a fit was obtained for a suppression factor 0.44 for the NLO calculation
with $x_{\gamma}^{obs} < 0.9$, based on fitting the distributions for 
$x_{\gamma}^{obs}$, $W$, $\bar{\eta}^{jets}$ and $E_T^{jet1}$. In this
comparison
they found disagreement for the largest $x_{\gamma}^{obs}$-bin and the lowest 
$\bar{\eta}^{jets}$ (which are related), but better agreement in the 
$E_T^{jet1}$-distribution.
In \cite{25} this leads to the statement, that the assumption that the
direct cross section obeys factorization is strongly disfavored by their
analysis. In total, it is obvious that in the final H1 analysis \cite{25}
a global suppression in diffractive dijet photoproduction is clearly
established and the model with resolved suppression only is not as well 
supported by the data.

Just recently also the ZEUS collaboration presented their final result on
diffractive dijet photoproduction \cite{27}. As in their preliminary
analysis, the two jets with the highest transverse energies $E_T^{jet}$ were
required to satisfy $E_T^{jet1(2)} > 7.5$
$(6.5)$ GeV, which is higher than in the
H1 analysis with $E_T^{jet1(2)} > 5$ $(4)$ GeV \cite{25}. ZEUS compared their 
measurements with the NLO predictions for diffractive photoproduction of dijets
based on our program \cite{21}. Three sets of DPDFs were used, the ZEUS LPS
fit, determined from a NLO analysis of inclusive diffraction and diffractive 
charm-production data \cite{3}, and the two H1 fits, H1 2006 fit A and fit B 
\cite{4}. The NLO results obtained with the two H1 fits were scaled down by a 
factor of 0.87 \cite{4}, since the H1 measurements used to derive the DPDFs 
include low-mass proton dissociative processes with $M_Y < 1.6$ GeV, 
which increases the photon-diffractive cross section by $1.15^{+0.15}_{-0.08}$
as compared to the pure proton final state as corrected to in the ZEUS
analysis. The comparison of the measured cross sections and the
theoretical predictions was based on the distributions in
the variables $y$, $M_X$, $x_{\p}$, $z_{\p}^{obs}$, $E_T^{jet1}$,
$\eta^{jet1}_{lab}$
and  $x_{\gamma}^{obs}$. The data were reasonably well described in their shape
as a function of these variables and lay systematically below the
predictions. The predictions for the three DPDFs differed appreciably. The
cross sections for the H1 2006 fit A (fit B) were the highest (lowest) and the
one for the ZEUS LPS fit lay between the two, but nearer to the fit A than the
fit B predictions. For $d\sigma/dx_{\gamma}^{obs}$ ZEUS also showed the ratio
of the data and the NLO predictions using the ZEUS LPS fit. It was 
consistent with a suppression factor of 0.7 independent of $x_{\gamma}^{obs}$.
This suppression factor depended
on the DPDFs and ranged between 0.6 (H1 2006 fit A) and 0.9 (fit B). Taking into
account the scale dependence of the theoretical predictions the ratio was
outside the theoretical uncertainty for the ZEUS LPS fit and the H1 2006 fit A,
but not for fit B. In their conclusions the authors
of the ZEUS analysis \cite{27} made the statement that the NLO
calculations tend to overestimate the measured cross section, which would
mean that a suppression is present. Unfortunately, however, they continued,
that, within the large
uncertainties of the NLO calculations, the data were compatible with the QCD
calculations, i.e.\ with no suppression.

Such a statement clearly contradicts the result of the H1 collaboration
\cite{25} and casts doubts on the correctness of the H1 analysis. The
authors of \cite{27} attribute this discrepancy to the fact that the H1
measurements \cite{25} were carried out in a lower $E_T^{jet}$ and a higher
$x_{\p}$ range than those in the ZEUS study \cite{27}.
Besides the different $E_T^{jet}$ and $x_{\p}$  regions in \cite{25} and 
\cite{27}, the two measurements suffer also from different experimental cuts of 
some other variables, which makes it difficult to compare the two data sets 
directly (note also the lower center-of-mass energy for the H1 data). In 
addition the comparison with NLO theory in \cite{25} and \cite{27} was done
with two different programs \cite{19} versus \cite{26}, which, however agreed
quite well with each other \cite{25}.

The rather different conclusions concerning factorization breaking in 
diffractive dijet photoproduction calls for a new comparative study of the two
data sets in \cite{25} and \cite{27}. We have therefore
performed a new calculation of the NLO cross sections on the basis of our
earlier work \cite{21} with the new H1 2006 DPDFs and revised hadronic 
corrections as compared to \cite{20}, in order to see whether we can confirm
the very different conclusions achieved in the H1 \cite{25} and ZEUS \cite{27}
analyses. In the comparison with the new data sets we shall follow more or less
the same strategy as in our earlier work \cite{20,21}. We first calculate
the unsuppressed NLO cross sections including an error band based on the scale
variation and see how much and in which distribution the data points lie inside
or outside this error band. Then we determine a global suppression factor by
fitting the differential cross section $d\sigma/dE_T^{jet1}$ at the bin with
the lowest $E_T^{jet1}$. With this suppression factor we shall compare to the
differential cross sections of all the other measured variables and look for
consistency. In this new comparison between the experimental and the 
theoretical results we shall concentrate on using the H1 2006 fit B \cite{4} 
input, since it leads to smaller cross sections than the DPDFs from H1 2006 
fit A \cite{4} or the ZEUS LPS DPDF fit \cite{3}.

Actually the H1 collaboration constructed a third set of DPDFs, which is called
the 'H1 2007 fit jets'. This fit is obtained through a simultaneous fit to the 
diffractive inclusive and DIS dijet cross sections \cite{28}. It is performed
under the assumption that there is no factorization breaking in the diffractive
dijet cross sections. Under this assumption, including the diffractive dijet
cross sections in the analysis leads to additional constraints, mostly on the
diffractive gluon distribution. On average the 'H1 2007 fit jets' is
similar to the 'H1 2006 fit B' except for the gluon distribution at high momentum
fraction and smaller factorization scales. In our analysis we shall 
disregard this new set of DPDFs, since it would be compatible with the
factorization test of the photoproduction data only if we restricted
these tests to the case that only the resolved part has this breaking and not
the direct part, which has the same structure as the DIS dijet cross section.

In Sec.\ 2 we shall present the complete list of cuts on the
experimental variables, give all the input used in the cross section
calculations, and present the basic formul\ae{}, from which the dijet cross 
sections have been calculated.  The comparison with the H1 \cite{25} and the
ZEUS \cite{27} experimental data is presented and discussed in Sec.\ 3.
In this comparison we shall concentrate on the main question, whether
there is a suppression in the photoproduction data at all. In addition we
shall investigate also whether a reasonable description of the data is
possible with suppression of the resolved cross section only, as we
studied it already in our previous work in 2004 \cite{20,21}. In
Sec.\ 4 we shall finish with a summary and our conclusions.

\section{Kinematic variables and cross section formul\ae{}}
\subsection{Kinematic variables and constraints}

The diffractive process $ep \rightarrow eXY$, in which the systems $X$ and 
$Y$ are separated by the largest rapidity gap in the final state, is
sketched in Fig.\ \ref{fig:1}.
%
%
The system $X$ contains at least two jets, and the system $Y$ is 
supposed to be a proton or another low-mass baryonic system. Let $k$ and $P$
denote the momenta of the incoming electron (or positron) and proton, 
respectively, and $q$ the momentum of the virtual photon $\gamma ^{*}$. Then 
the usual kinematic variables are
\begin{equation}
 s = (k+P)^2,~ Q^2 = -q^2,~{\rm and}~y = \frac{qP}{kP}.
\end{equation}
We denote the four-momenta of the systems $X$ and $Y$ by $p_X$ and $p_Y$. 
The H1 data \cite{25} are described in terms of
\begin{eqnarray}
 M_X^2 = p_X^2 & ~{\rm and}~ & t = (P-p_Y)^2, \nonumber\\
 M_Y^2 = p_Y^2 & ~{\rm and}~ & x_{\p} = \frac{q(P-p_Y)}{qP},
\end{eqnarray}
where $M_X$ and $M_Y$ are the invariant masses of the systems $X$ and $Y$,
$t$ is the squared four-momentum transfer of the incoming proton and the
system $Y$, and $x_{\p}$ is the momentum fraction of the proton beam
transferred to the system $X$.

\begin{table}
 \tbl{\label{tab:1}Kinematic cuts applied in the H1 analysis of
 diffractive dijet photoproduction.}{
 \begin{tabular}{rcccl}
 \hline\noalign{\smallskip}
  165 GeV  & $<$ & $W$   & $<$ & 242 GeV \\
        &     & $Q^2$ & $<$ & 0.01 GeV$^2$ \\
        &     & $E_T^{jet1}$ & $>$ & 5 GeV \\
        &     & $E_T^{jet2}$ & $>$ & 4 GeV \\
  $-1$  & $<$ & $\eta_{lab}^{jet1,2}$ & $<$ & 2 \\
        &     & $x_{\p}$ & $<$ & 0.03 \\
        &     & $M_Y$ & $<$ & 1.6 GeV \\
        &     & $-t$  & $<$ & 1 GeV$^2$ \\
 \noalign{\smallskip}\hline
 \end{tabular}}
\end{table}
The exchange between the systems $X$ and $Y$ is supposed to be the pomeron 
$\p$ or any other Regge pole, which couples to the proton and the system
$Y$ with four-momentum $P-p_Y$. In this work we will neglect Reggeon exchanges,
which contribute only at large $x_{\p}$.
The pomeron is resolved into partons (quarks
or gluons) with four-momentum $v$. In the same way the virtual photon can
resolve into partons with four-momentum $u$, which is equal to $q$ for the
direct process. With these two momenta $u$ and $v$ we define
\begin{equation}
 x_{\gamma } = \frac{Pu}{Pq} ~{\rm and}~ z_{\p} = \frac{qv}{q(P-p_Y)}.
\end{equation}
$x_\gamma$ is the longitudinal-momentum fraction carried by the partons 
coming from the photon, and $z_{\p}$ is the corresponding quantity carried
by the partons of the pomeron etc., i.e.\ the diffractive exchange. For
the direct process we have $x_{\gamma } = 1$. The final state, produced by
the ingoing momenta $u$ and $v$, has the invariant mass $M_{12} =
\sqrt{(u+v)^2}$, which is equal to the invariant dijet mass in the case that
no more than two hard jets are produced. $q-u$ and $P-p_Y-v$ are the
four-momenta of the remnant jets produced at the photon and pomeron side.
The regions of the kinematic variables, in which the cross section has been
measured by the H1 collaboration \cite{25}, are given in Tab.~\ref{tab:1},
whereas the corresponding regions for the ZEUS analysis \cite{27}
are given in Tab.\ 2.
In each case, we have evaluated the theoretical cross sections with the
corresponding constraints.

The upper limit of $x_{\p}$ is kept small in order for the pomeron exchange
to be dominant. In the experimental analysis as well as in the NLO
calculations, jets are defined with the inclusive $k_T$-cluster algorithm
with a distance parameter $d=1$ \cite{29} in the laboratory frame. At least
two jets are required with transverse energies $E_T^{jet1} > 5$ (7.5) GeV and 
$E_T^{jet2} > 4$ (6.5) GeV. They are the leading and subleading jets with 
$-1<\eta_{lab}^{jet1,2} < 2$ ($-1.5<\eta_{lab}^{jet1,2} < 1.5$) for H1 (ZEUS).
The lower limits of the jet $E_T$'s
are asymmetric in order to avoid infrared sensitivity in the computation of
the NLO cross sections, which are integrated over $E_T$ \cite{Klasen:1995xe}.

In the experimental analysis the variable $y$ is deduced from the energy 
$E_e'$ of the scattered electron, $y=1-E_e'/E_e$. Furthermore,
$sy=W^2=(q+P)^2=(p_X+p_Y)^2$. The range of $W$ given in Tab.\ 1 corresponds 
to the $y$ range $0.3<y<0.65$. $x_{\p}$ is reconstructed according to
\begin{equation}
 x_{\p} = \frac{\sum_{X} (E+p_z)}{2E_p},
\end{equation}
where $E_p$ is the proton beam energy and the sum runs over all particles
(jets) in the $X$-system. The variables $M_{12}$, $x_{\gamma }$, and
$z_{\p}$ are determined only from the kinematic variables of the two hard
leading jets with four-momenta $p^{jet1}$ and $p^{jet2}$. So,
\begin{equation}
 M_{12}^2 = (p^{\rm jet1} + p^{\rm jet2})^2,
\end{equation}
where additional jets are not taken into account. In the same way we have
\begin{equation}
 x_{\gamma }^{\rm obs} =\frac{\sum_{\rm jets}(E-p_z)}{2yE_e} ~{\rm and}~
 z_{\p}^{\rm obs} = \frac{\sum_{\rm jets}(E+p_z)}{2x_{\p} E_p}.
\end{equation}
The sum over jets runs only over the variables of the two leading jets.
These definitions for $x_{\gamma }$ and $z_{\p}$ are not the same as the
definitions given earlier, where also the remnant jets and any additional
hard jets are taken into account in the final state. In the same way $M_X$
can be estimated by $M_X^2 = M_{12}^2/(z_{\p}^{\rm obs} x_{\gamma}^{\rm
obs})$. The dijet system is characterized by the transverse energies
$E_T^{jet1}$ and $E_T^{jet2}$ and the rapidities in the laboratory
system $\eta^{jet1}_{lab}$ and $\eta^{jet2}_{lab}$. The
differential cross sections are measured and calculated as functions of the
transverse energy $E_T^{jet1}$ of the leading jet, the average rapidity
$\bar{\eta}^{jets} = (\eta^{jet1}_{lab} + \eta^{jet2}_{lab})/2$,
and the jet separation $|\Delta\eta^{jets}| =
|\eta^{jet1}_{lab} - \eta^{jet2}_{lab}|$, which is related
%
\begin{table}
 \tbl{Kinematic cuts applied in the ZEUS analysis of
 diffractive dijet photoproduction.}{
 \begin{tabular}{rcccl}
 \hline
~~~  0.2~~~ & $<$ ~~& $y$    ~~& $<$ & 0.85     \\
      &     & ~~$Q^2$        ~~& $<$ & 1 GeV$^2$  \\
      &     & ~~$E_T^{jet1}$ ~~& $>$ & 7.5 GeV \\
      &     & ~~$E_T^{jet2}$ ~~& $>$ & 6.5 GeV \\
~~~ -1.5~~~ & $<$ ~~& $\eta^{jet1,2}_{lab}$ ~~& $<$ & 1.5 \\
      &     & ~~$x_{\p}$     ~~& $<$ & 0.025 \\
      &     & ~~$-t$         ~~& $<$ & 5 GeV$^2$ \\
 \noalign{\smallskip}\hline
 \end{tabular}}
\end{table}
%
to the scattering angle in the center-of-mass system of the two jets.

\subsection{Diffractive parton distributions}
The diffractive PDFs are obtained from an analysis of the diffractive
process $ep \rightarrow eXY$, which is illustrated in Fig.\ \ref{fig:0},
where now $Q^2$ is large and the state $X$ consists of all possible final
states, which are summed. The cross section for this diffractive DIS process
depends in general on five independent variables (azimuthal angle dependence
neglected): $Q^2$, $x$ (or $\beta$), $x_{\p}$, $M_Y$, and $t$. These
variables are defined as before, and $x= Q^2/(2Pq) = Q^2/(Q^2+W^2) = x_{\p}
\beta$. The system $Y$ is not measured, and the results are integrated over
$-t < 1$ GeV$^2$ and $M_Y < 1.6$ GeV as in the photoproduction case. The
measured cross section is expressed in terms of a reduced diffractive cross
section $\sigma^{D(3)}_r$ defined through
\begin{equation}
 \frac{d^3 \sigma^D}{dx_{\p}dxdQ^2} = \frac{4\pi \alpha^2}{xQ^4}
 \left(1-y+\frac{y^2}{2}\right) \sigma^{D(3)}_r(x_{\p},x,Q^2)
\end{equation}
and is related to the diffractive structure functions $F^{D(3)}_2$ and
$F^{D(3)}_L$ by
\begin{equation}
 \sigma^{D(3)}_r = F^{D(3)}_2 - \frac{y^2}{1+(1-y)^2} F^{D(3)}_L.
\end{equation}
$y$ is defined as before, and $F^{D(3)}_L$ is the longitudinal diffractive
structure function.

The proof of Collins \cite{10}, that QCD factorization is applicable to
diffractive DIS, has the consequence that the DIS cross section for 
$\gamma ^{*}p \rightarrow XY$ can be written as a convolution of a partonic 
cross section $\sigma^{\gamma ^{*}}_a$, which is calculable as an expansion 
in the strong coupling constant $\alpha_s$, with diffractive PDFs $f^D_a$ 
yielding the probability distribution for a parton $a$ in the proton under
the constraint that the proton undergoes a scattering with a particular
value for the squared momentum transfer $t$ and $x_{\p}$. Then the cross
section for $\gamma^{*} p \rightarrow XY$ is
\begin{equation} 
 \frac{d^2\sigma }{dx_{\p}dt} = \sum_{a} \int_{x}^{x_{\p}} d\xi
 \sigma^{\gamma *}_{a}(x,Q^2,\xi) f_a^D(\xi,Q^2;x_{\p},t).
\end{equation}
This formula is valid for sufficiently large $Q^2$ and fixed $x_{\p}$ and
$t$. The parton cross sections are the same as those for inclusive DIS. The
diffractive PDFs are non-perturba\-tive objects. Only their $Q^2$-evolution
can be predicted with the well-known DGLAP evolution equations \cite{2}, 
which we shall use in NLO.

Usually for $f_a^D(x,Q^2;x_{\p},t)$ an additional assumption is made, namely
that it can be written as a product of two factors, $f_{\p/p}(x_{\p},t)$ and
$f_{a/\p}(\beta,Q^2)$,
\begin{equation}
 f_a^D(x,Q^2;x_{\p},t) = f_{\p/p}(x_{\p},t) f_{a/\p}(\beta=x/x_{\p},Q^2).
\end{equation}
$f_{\p/p}(x_{\p},t)$ is the pomeron flux factor. It gives the probability
that a pomeron with variables $x_{\p}$ and $t$ couples to the proton. Its
shape is controlled by Regge asymptotics and is in principle measurable by
soft processes under the condition that they can be fully described by
single-pomeron exchange. This Regge factorization formula
represents the resolved pomeron model, in
which the diffractive exchange, i.e.\ the pomeron, can be considered as
a quasi-real particle with a partonic structure given by PDFs $f_{a/\p}
(\beta,Q^2)$. $\beta$ is the longitudinal momentum fraction of the pomeron
carried by the emitted parton $a$ in the pomeron. The important point is
that the dependence of $f_a^D$ on the four variables $x, Q^2, x_{\p}$ and
$t$ factorizes into two functions $f_{\p/p}$ and $f_{a/\p}$, which each depend
only on two variables.

Since the value of $t$ could not be fixed in the diffractive DIS
measurements, it is integrated over with $t$ in the region
$t_{\rm cut} < t < t_{\min}$. Therefore we have \cite{4,12}
\begin{equation}
 f(x_{\p}) = \int_{t_{\rm cut}}^{t_{\min}} dt f_{\p/p}(x_{\p},t),
\end{equation}
where $t_{\rm cut} = -1$ GeV$^2$ and $t_{\min}$ is the minimum kinematically
allowed value of $|t|$. In \cite{12,4} the pomeron flux factor is assumed to
have the form
\begin{equation}
 f_{\p/p}(x_{\p},t) = x_{\p}^{1-2\alpha_ {\p}(t)} \exp (B_{\p} t).
 \label{eq:12}
\end{equation}
$\alpha _{\p}(t)$ is the pomeron trajectory, $\alpha _{\p}(t)=\alpha _{\p}
(0) + \alpha'_{\p} t$, assumed to be linear in $t$. The values of $B_{\p},
\alpha _{\p}(0)$ and $\alpha _{\p}'$ are taken from \cite{4} and have the
values $B_{\p}=5.5$ GeV$^{-2}$, $\alpha_{\p}(0)=1.118$ (fit A),
$\alpha_{\p}(0)=1.111$ (fit B) and $\alpha_{\p}'= 0.06$ GeV$^{-2}$. 
Usually $f_{\p/p}(x_{\p},t)$ as written in Eq.\
(\ref{eq:12}) has in addition to the dependence on $x_{\p}$ and $t$ a
normalization factor $N$, which can be inferred from the asymptotic behavior
of $\sigma_{\rm tot}$  for $pp$  and $p\bar{p}$ scattering. Since it is
unclear whether these soft diffractive cross sections are dominated by a
single pomeron exchange, it is better to include $N$ into the pomeron PDFs
$f_{a/\p}$ and fix it from the diffractive DIS data \cite{4}. The
diffractive DIS cross section $\sigma _r^{D(3)}$ is measured in the
kinematic range $3.5 \leq Q^2 \leq 1600$ GeV$^2$, $0.01 \leq \beta \leq 0.9$
and $10^{-4} \leq x_{\p} < 0.05$.

The pomeron couples to quarks in terms of a light flavor singlet $\Sigma
(z_{\p})=u(z_{\p})+d(z_{\p})+s(z_{\p})+\bar{u}(z_{\p})+\bar{d}(z_{\p})+
\bar{s}(z_{\p})$ and to gluons in terms of $g(z_{\p})$, which are
parameterized at the starting scale $Q_0^2=1.75$ GeV$^2$ (fit A) and $2.5$ 
GeV$^2$ (fit B). $z_{\p}$ is the momentum fraction entering the hard 
subprocess, so that for the LO process
$z_{\p}=\beta$, and in NLO $\beta < z_{\p} < 1$. These PDFs of the pomeron
are parameterized by a particular form in terms of the usual power ansatz as
given in \cite{4}. Charm quarks and bottom quarks couple differently from the 
light quarks by including the finite charm mass $m_c=1.4$ GeV and bottom mass
$m_b=4.5$ GeV in the massive quark scheme and describing the coupling to 
photons via the photon-gluon fusion type process up to order $\alpha_s^2$.
For the pomeron PDFs, we used a two-dimensional fit in the variables
$z_{\p}$ and $Q^2$ and then inserted the interpolated result in the cross
section formula.

\subsection{Cross section formula}
Under the assumption that the cross section can be calculated from the well-known
formul\ae{} for jet production in low-$Q^2$ $ep$ collisions, the cross
section for the reaction $e+p \rightarrow e+2~{\rm jets}+X'+Y$ is computed
from the following basic formula:
\begin{eqnarray}
 && d\sigma ^D(ep \rightarrow e+2~{\rm jets}+X'+Y) = \nonumber \\ 
 && \sum_{a,b} \int_{t_{\rm cut}}^{t_{\min}}dt \int_{x_{\p}^{\min}}^{x_{\p}
  ^{\max}} dx_{\p} \int_{0}^{1}dz_{\p} \int_{y_{\min}}^{y_{\max}}dy \int_{0}
  ^{1}dx_{\gamma} \nonumber \\
 && f_{\gamma/e}(y) f_{a/\gamma }(x_\gamma,M^2_{\gamma }) f_{\p/p}(x_{\p},t)
  f_{b/\p}(z_{\p},M_{\p}^2) \nonumber \\
 && d\sigma^{(n)} (ab \rightarrow {\rm jets}).
 \label{eq:13}
\end{eqnarray}
$y$, $x_\gamma$ and $z_{\p}$ denote the longitudinal momentum fractions of
the photon in the electron, the parton $a$ in the photon, and the parton $b$
in the pomeron. $M_{\gamma}$ and $M_{\p}$ are the factorization scales at
the respective vertices, and $d\sigma ^{(n)}(ab \rightarrow {\rm jets)}$ is
the cross section for the production of an $n$-parton final state from two
initial partons $a$ and $b$. It is calculated in NLO, as are the
PDFs of the photon and the pomeron.

The function $f_{\gamma /e}(y)$, which describes the virtual photon
spectrum, is assumed to be given by the well-known Weizs\"acker--Williams
approximation,
 \begin{eqnarray}
 f_{\gamma /e}(y) &=& \frac{\alpha }{2\pi} \left[\frac{1+(1-y)^2}{y} \ln 
\frac{Q^2_{\max}(1-y)}{m_e^2y^2} \right. \nonumber \\
 &+& \left. 2m_e^2y\lr\frac{1-y}{m_e^2y^2} -\frac{1}{Q^2_{\max}}\rr\right].
\end{eqnarray}
Usually, only the dominant leading logarithmic contribution is considered.
We have added the second non-logarith\-mic term as evaluated in \cite{31}.
$Q^2_{\max} = 0.01$ (1) GeV$^2$ for the H1 (ZEUS) cross sections calculated
in this work.

The formula for the cross section $d\sigma ^D$ can be used for the resolved
as well as for the direct process. For the latter, the parton $a$ is the
photon and $f_{\gamma /\gamma }(x_\gamma,M^2_{\gamma})=\delta(1-x_\gamma)$,
which does not depend on $M_{\gamma }$. As is well known, the distinction
between direct and resolved photon processes is meaningful only in LO of
perturbation theory. In NLO, collinear singularities arise from the photon
initial state, that must be absorbed into the photon PDFs and produce a
factorization scheme dependence as in the proton and pomeron cases. The
separation between the direct and resolved processes is an artifact of
finite order perturbation theory and depends in NLO on the factorization
scheme and scale $M_{\gamma}$. The sum of both parts is the only physically
relevant quantity, which is approximately independent of the factorization
scale $M_{\gamma}$ due to the compensation of the scale dependence between
the NLO direct and the LO resolved contribution.
For the resolved process, PDFs of the photon are need\-ed, for which we
choose the NLO versions of GRV \cite{32} transformed to the $\overline{\rm MS}$
scheme.

\section{Results}
\subsection{Comparison with H1 data}

In this section, we present the comparison of the various theoretical 
predictions in NLO with the experimental data from the H1 collaboration 
\cite{25}.  The corresponding kinematic cuts are given in Tab.\ 1. 
Before we confront the calculated cross sections with the experimental
data, we correct them for hadronization effects. The hadronization
corrections are calculated by means of the LO RAPGAP Monte Carlo generator 
\cite{17}. The factors for the transformation from jets made up of stable
hadrons to parton jets were supplied by the H1 collaboration \cite{25}. Most 
of our calculations are done with the `H1 2006 fit B' \cite{4} DPDFs since
they 
give they smaller diffractive dijet cross sections as compared to the 
`H1 2006 fit A'. These DPDF fits are based on $n_f=3$ massless flavors. The 
production of charm and bottom quarks was treated there in the Fixed-Flavor 
Number Scheme (FFNS) in NLO with non-zero charm and bottom quark mass. Instead 
of this extra treatment of the charm and bottom contribution in the pomeron 
we added a charm PDF in the pomeron as obtained in the `H1 2002 fit' \cite{12},
where the charm quark was also considered to be massless. The bottom 
contribution was neglected. This assumption simplifies the calculations
considerably.  Since the
charm contribution from the pomeron is small, this should be a good
approximation.
We then take $n_f=4$ with  
$\Lambda^{(4)}_{\overline{\rm MS}} = 0.347$ GeV, which corresponds to the value
used in the DPDFs `H1 2006 fit A' and `H1 2006 fit B' \cite{4}.

As it is clear from the discussion of the various preliminary analyses of
the H1 and ZEUS collaborations, there are two questions which we would
like to answer from the comparison with the recent H1 and the ZEUS data.
The first question is whether a suppression, which differs substantially from 
one, is needed to describe the data. The second question is whether the data 
are also consistent with a suppression factor applied to the resolved cross 
section only. To give an answer to these two questions we calculated first the 
cross sections with no suppression factor ($R = 1$ in the following figures) 
with a theoretical error obtained from varying the common scale of 
renormalization and factorization by factors of 0.5  and 2 around the central 
scale (highest $E_T^{jet}$). In a second step we show the results  
for the same differential cross sections with a global suppression factor,
adjusted to $d\sigma/dE_T^{jet1}$ in the smallest $E_T^{jet1}$-bin. As in the
experimental analysis \cite{25}, we consider the differential cross sections
in  the variables $x_{\gamma}^{obs}$, $z_{\p}^{obs}$, $\log_{10}(x_{\p})$,
$E_T^{jet1}$,  $M_{12}$, $\bar{\eta}^{jets}$, $|\Delta \eta^{jets}|$ 
and $W$.

The unsuppressed ($R=1$) cross sections $d\sigma/dx_{\gamma}^{obs}$,
$d\sigma/dz_{\p}^{obs}$, $d\sigma/d\log_{10}(x_{\p})$, $d\sigma/dE_T^{jet1}$,
$d\sigma/dM_{12}$, $d\sigma/d\bar{\eta}^{jets}$, $d\sigma/d|\Delta
\eta^{jets}|$
and $d\sigma/dW$ ($\bar{\eta}^{jets} \equiv \langle\eta_{lab}^{jet}\rangle$ in
\cite{25})
with their scale variation are shown in Fig.\ 3a-h.
In these figures we also plotted the experimental data with their errors. 
Except for two points (largest $z_{\p}^{obs}$ and largest $E_T^{jet1}$-bin) all
other experimental points lie, including their errors, outside the theoretical
error band. This comparison clearly tell us, that an unsuppressed cross section
is in disagreement with the data. It is clear, that with the DPDFs 'H1 2006
fit A'
cross section this conclusion would be even stronger, since with these DPDFs
the unsuppressed cross sections are even larger. That $d\sigma/dz_{\p}^{obs}$
overlaps in the largest bin with the lower limit of the prediction for $R=1$
(see Fig.\ 3b) can be explained with the fact that the gluon DPDF in the
'H1 2006 fit B' is not very well constrained for large $\beta$ and might be
larger  there.

If we now determine the suppression factor from fitting the lowest 
$E_T^{jet1}$-bin experimental cross section we obtain $R=0.42 \pm 0.06$.
The indicated error corresponds to the experimental uncertainty, while we
show in the figures explicitly the theoretical uncertainty. With 
this suppression factor we have calculated the eight distributions including 
their theoretical errors and compare with the experimental data including their
errors. The results of this comparison is shown also in Figs.\ 3a-h.
%
\begin{figure}
 \centering
 \includegraphics[width=0.325\columnwidth]{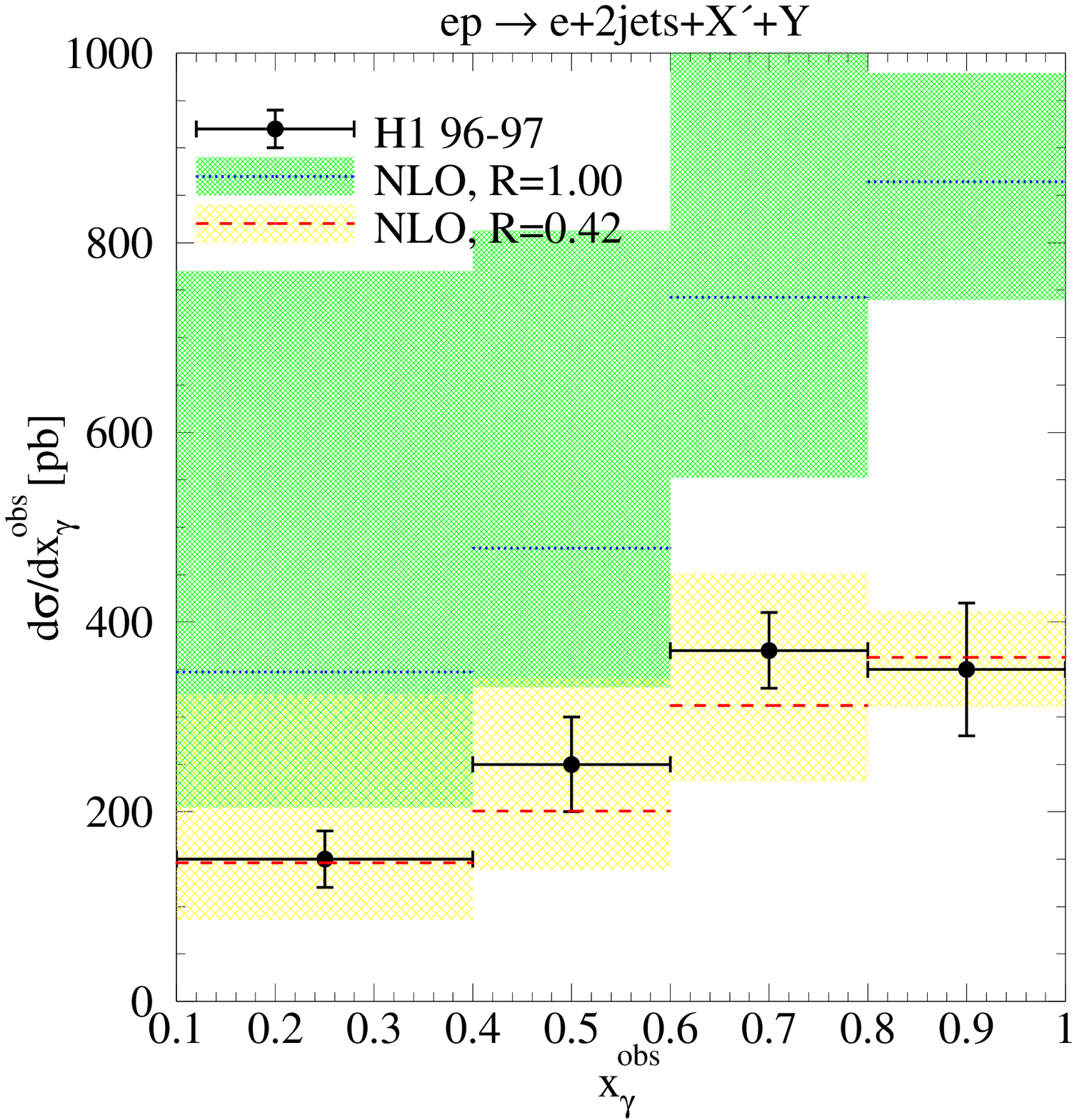}
 \includegraphics[width=0.325\columnwidth]{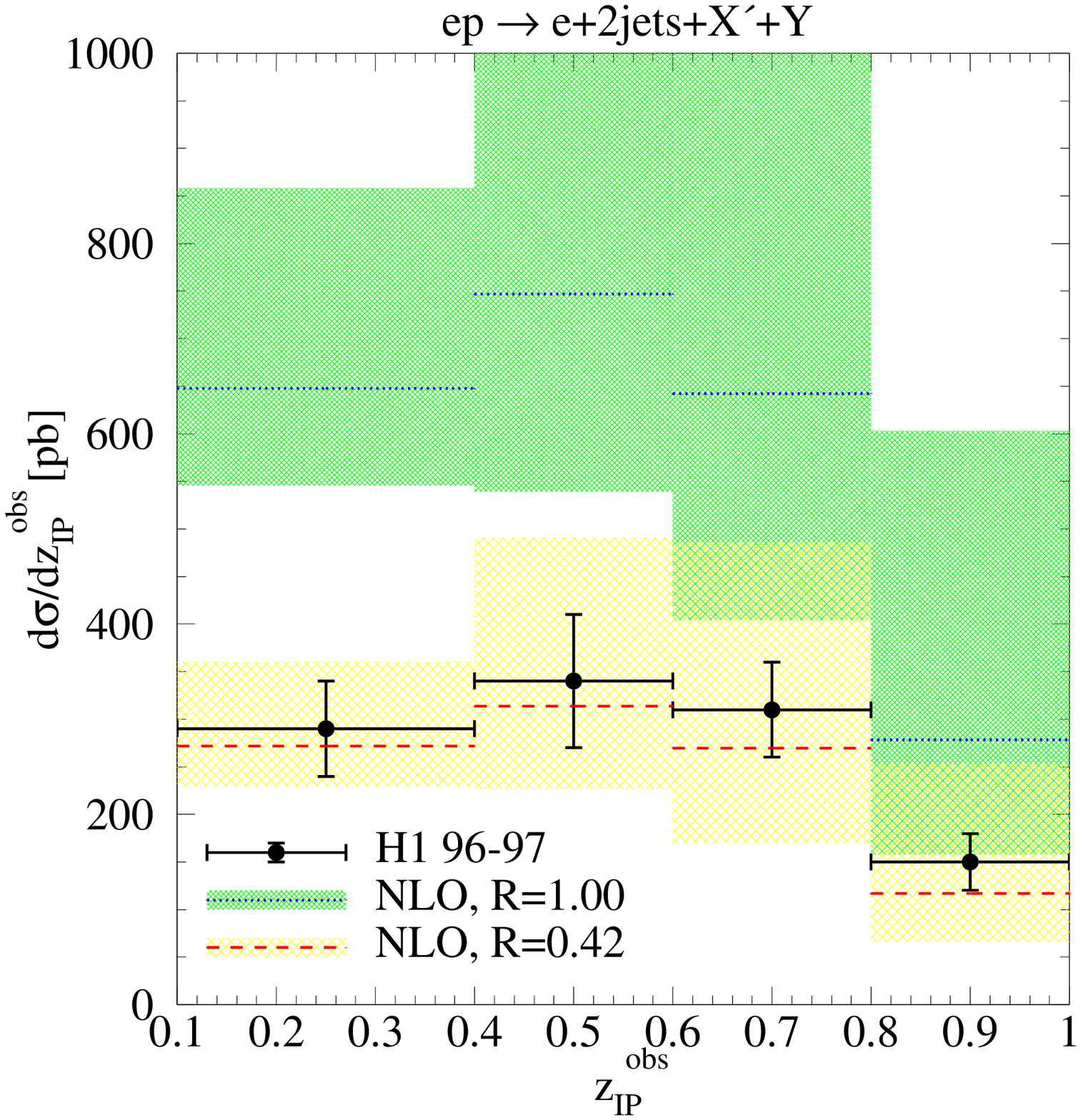}
 \includegraphics[width=0.325\columnwidth]{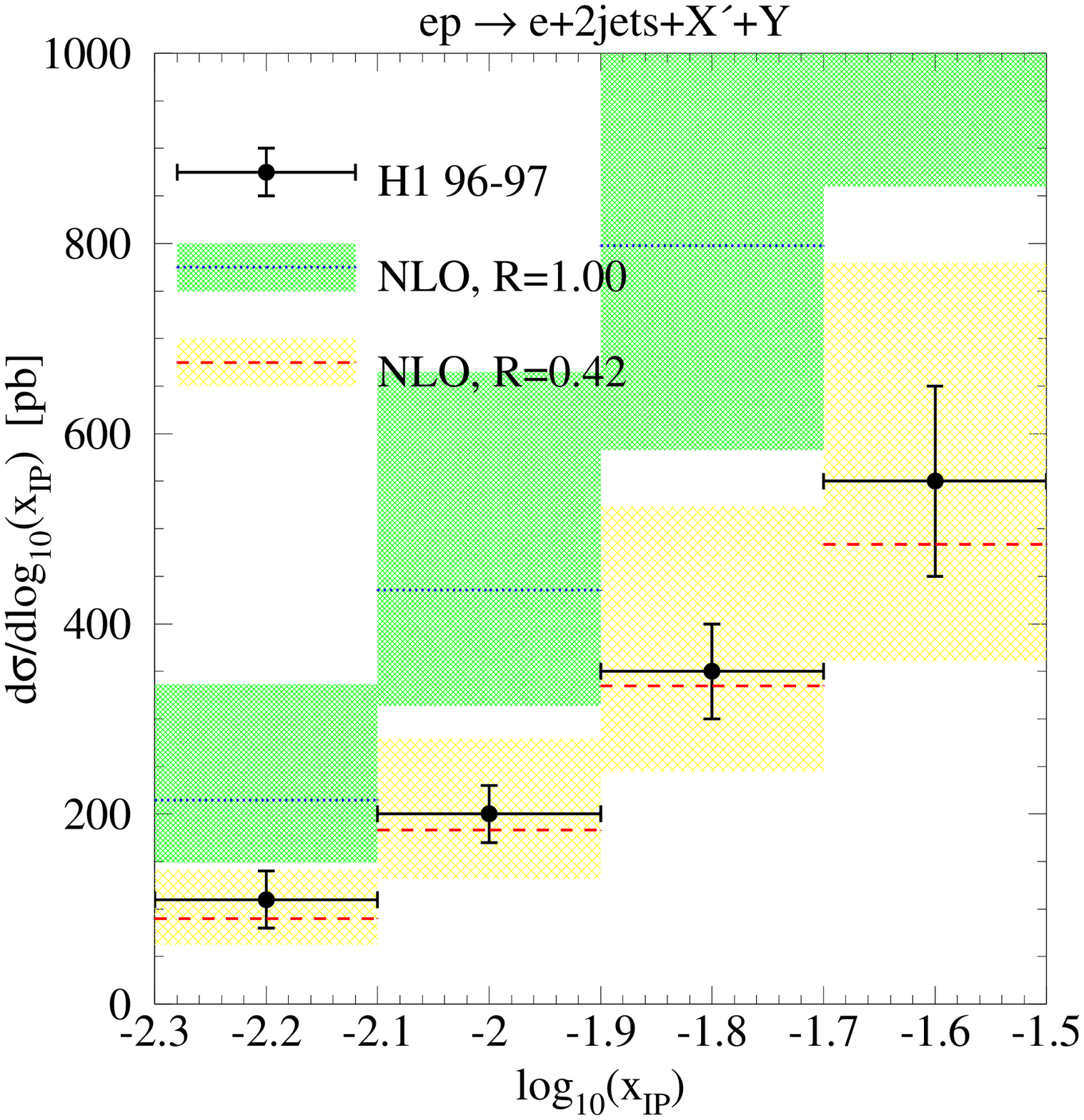}
 \includegraphics[width=0.325\columnwidth]{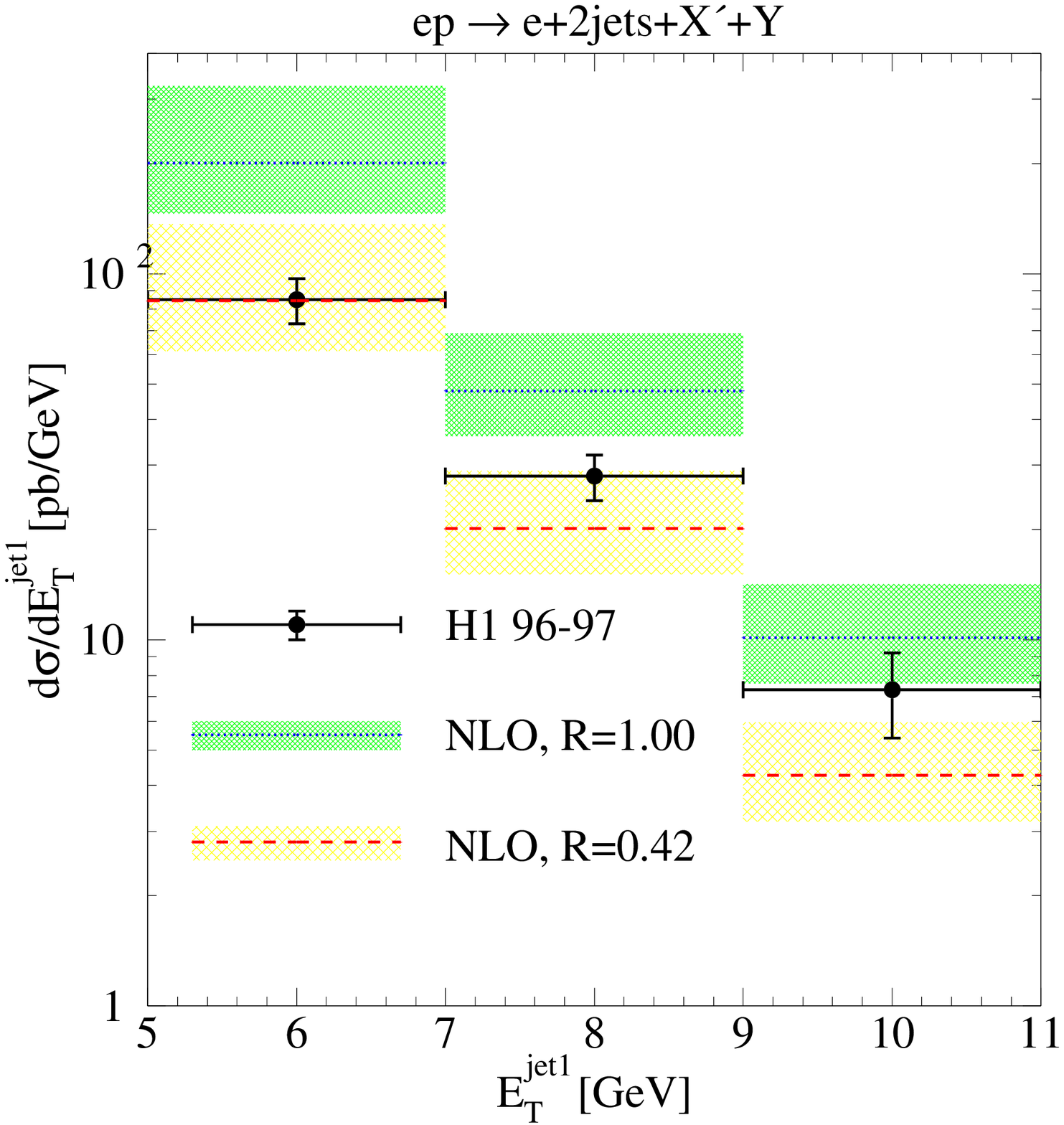}
 \includegraphics[width=0.325\columnwidth]{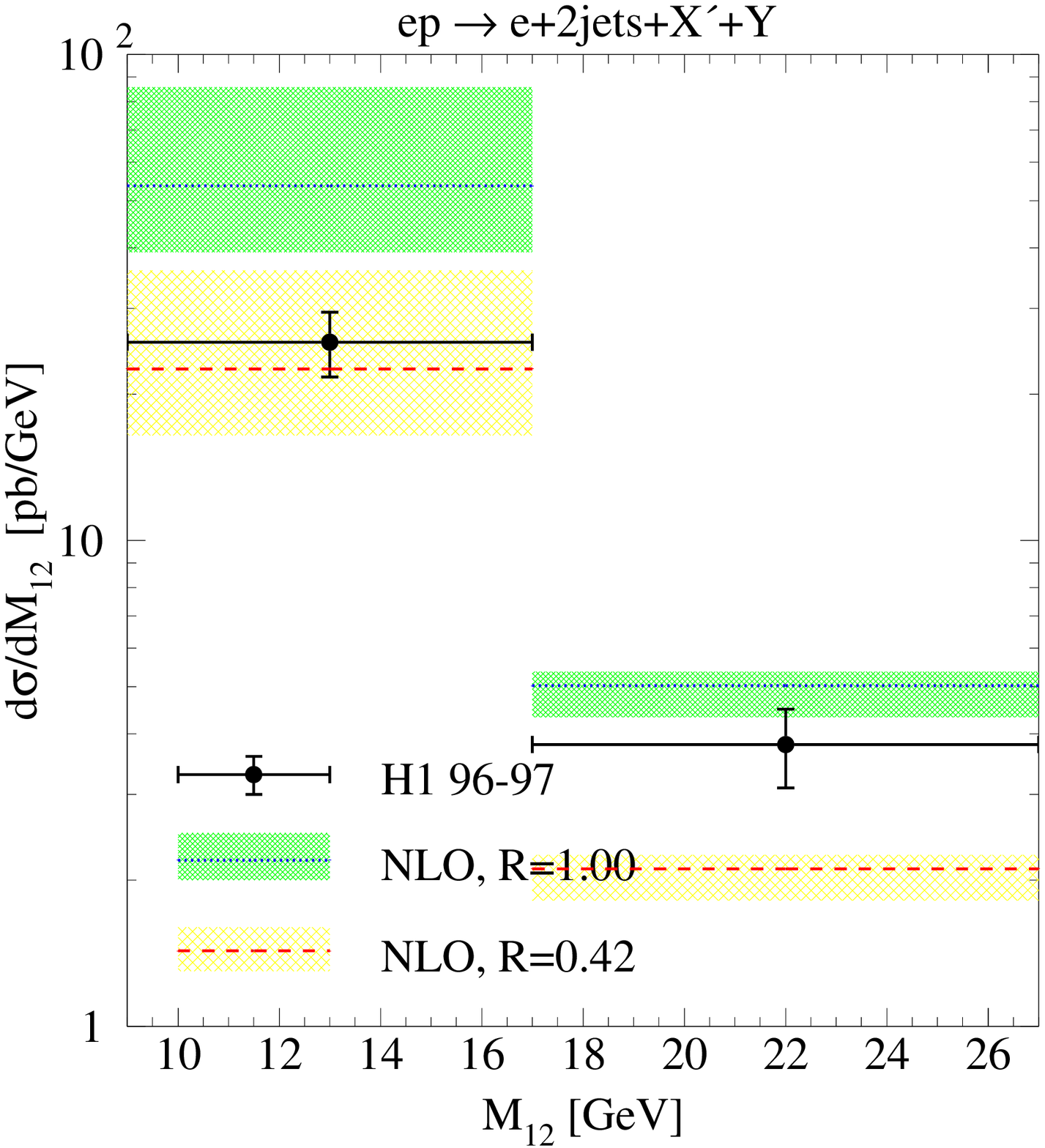}
 \includegraphics[width=0.325\columnwidth]{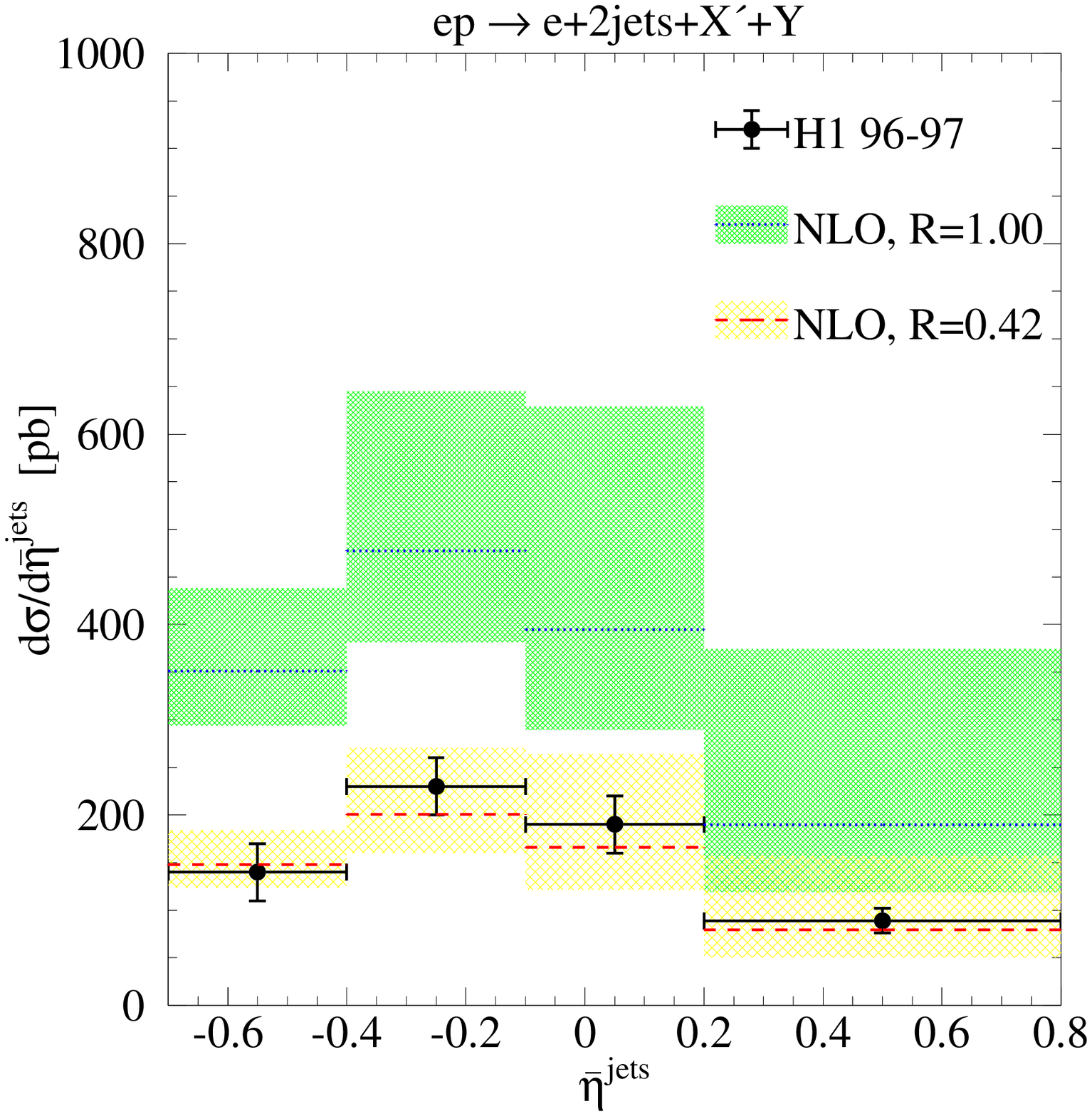}
 \includegraphics[width=0.325\columnwidth]{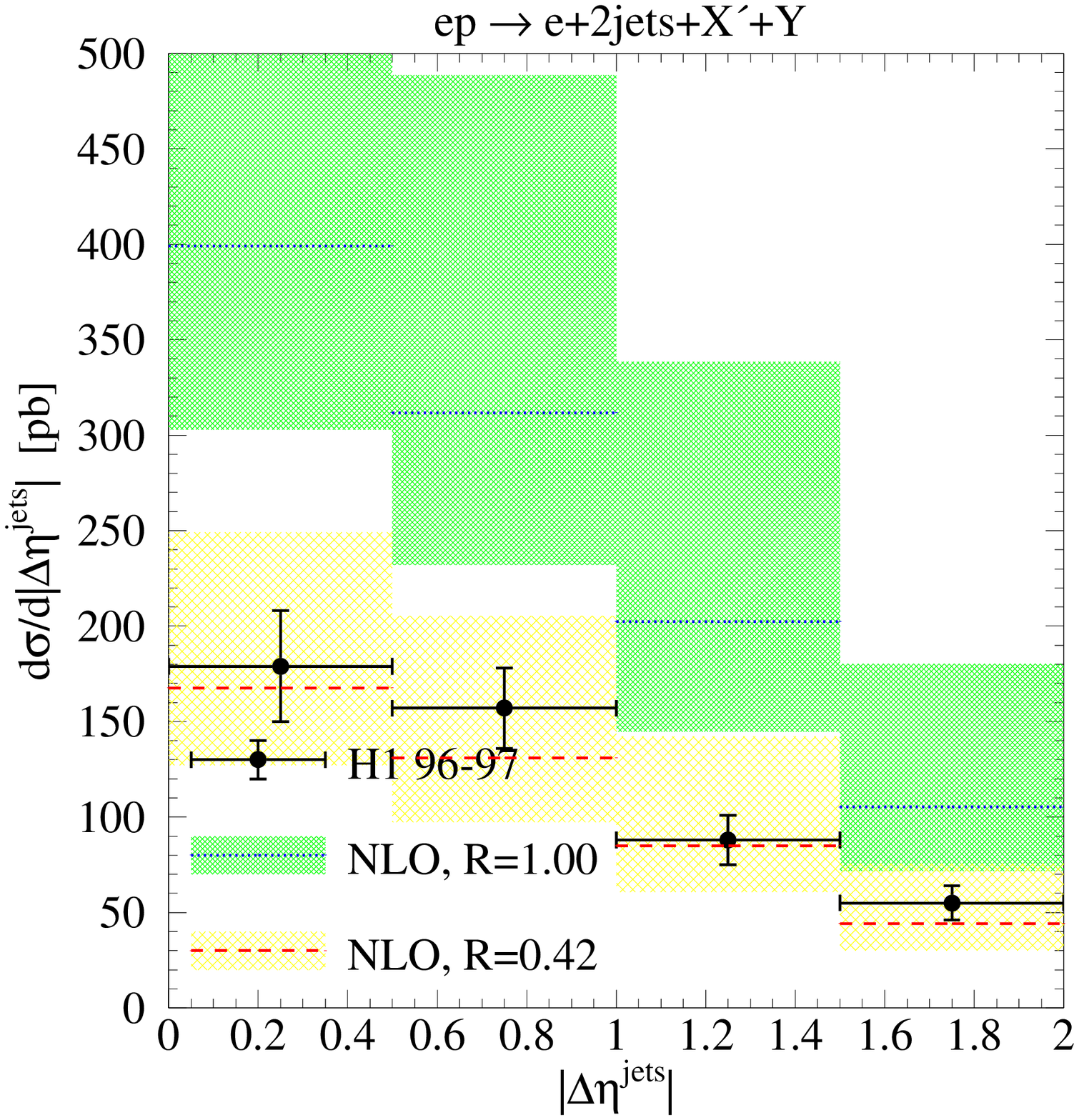}
 \includegraphics[width=0.325\columnwidth]{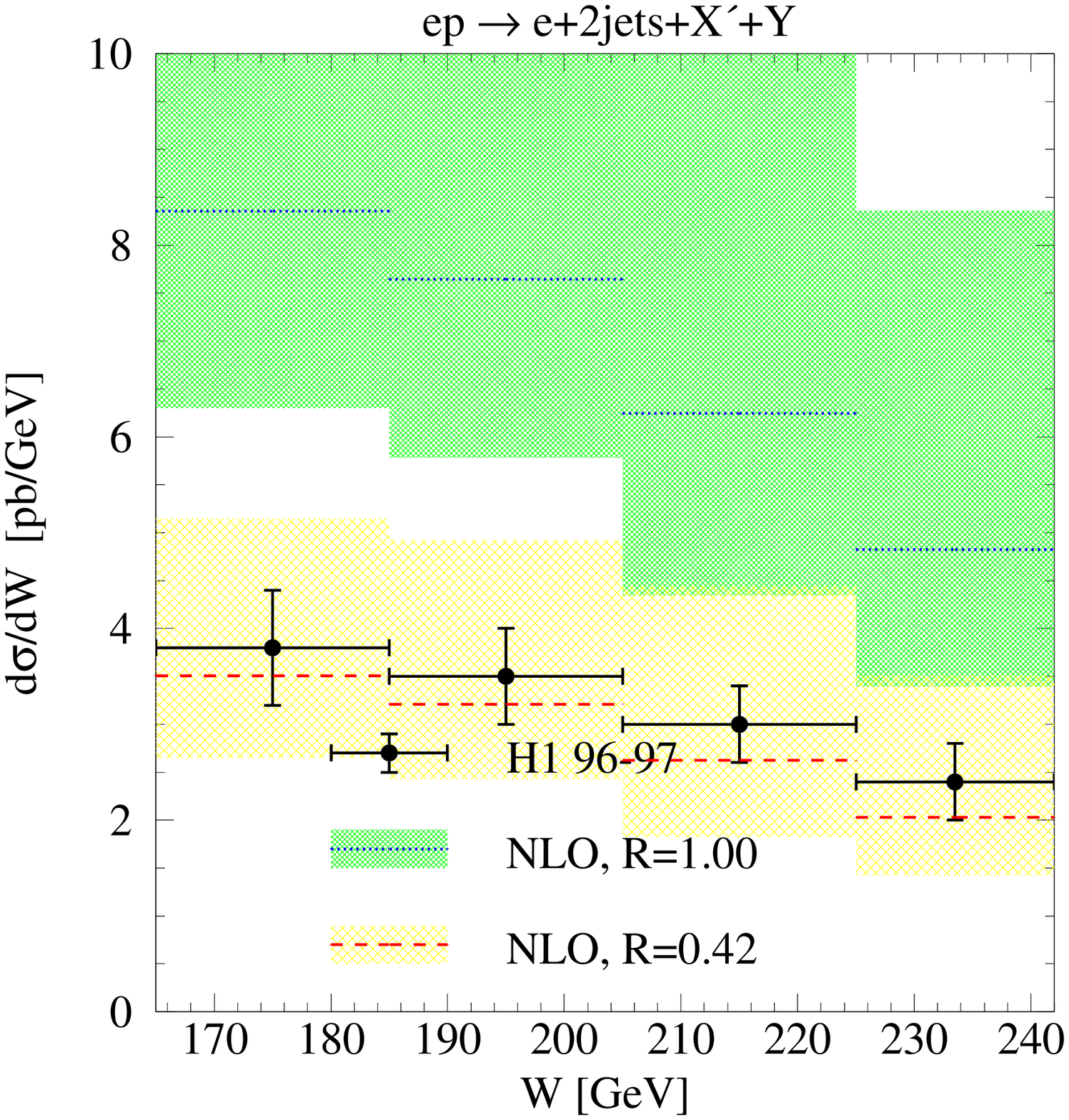}
 \caption{\label{fig:2}Differential cross sections for diffractive dijet
 photoproduction as measured by H1 and compared to
 NLO QCD without ($R=1$) and with ($R=0.42$) global suppression
 (color online).}
\end{figure}
%
With the exception of Figs.\ 3d and 3h, where the comparisons of 
$d\sigma/dE_T^{jet1}$ and $d\sigma/dM_{12}$ are shown, all other plots are 
such that the data points lie inside the error
band based on the scale variation. Most of the data points even agree with the 
$R = 0.42$ predictions inside the much smaller experimental errors. 
In $d\sigma/dE_T^{jet1}$  
(see Fig.\ 3d) the predictions for the second and third bin
lie outside the data points with their errors. For $R = 1$ and $R=0.42$
these cross sections falls off stronger with increasing $E_T^{jet1}$  than the 
data, the normalization being of course about two times larger for $R=1$.
In particular, the third data point agrees with the $R = 1$ prediction.
This means that the suppression decreases with increasing $E_T^{jet1}$. Such 
a behavior points in the direction that a suppression of the resolved cross 
section only would give better agreement with the data, as we shall see below. 
The same observations can be made by looking at $d\sigma/dM_{12}$ in Fig.\ 3e. 
The survival probability $R = 0.42\pm0.06$ agrees with the result in \cite{25},
which quotes $R=0.50 \pm 0.10$, determined by a fit to the double ratio of
measured to predicted cross section in photoproduction by the corresponding
ratio in DIS given as a function of $W$. In this double ratio many experimental
errors and theoretical scale errors cancel to a large extent. This double ratio
is also insensitive to the detailed shape of the diffractive gluon density.
From our comparison we conclude that the H1 data show a global suppression of 
the order of  two in complete agreement with the results \cite{20,21} and 
\cite{25} based on earlier preliminary \cite{16} and final H1 data \cite{25}.

Next we want to answer the second question, whether the data could be
consistent with a suppression of the resolved component only. For this
purpose we have calculated the cross sections in two versions: (i)
suppression of the resolved cross section and (ii)
suppression of the resolved cross section plus that part of the NLO direct
part which depends on the factorization scale  at the photon vertex
and thereby eliminates the $M_\gamma$-dependence in the combined direct and
resolved cross section
\cite{34}. Of course, the needed suppression factors for the two versions will 
be different. We determine the suppression factors by fitting again the
measured $d\sigma/dE_T^{jet1}$ for the lowest $E_T^{jet1}$-bin (see Fig.\
4d). Then, the suppression factor for version (i) is $R = 0.31$ (denoted res 
in the figures), and for version (ii) it is $R = 0.29$ (denoted res+dir-IS). 
The comparison with the H1 data of $d\sigma/dx_{\gamma}^{obs}$, 
$d\sigma/dz_{\p}^{obs}$, $d\sigma/d\log_{10}(x_{\p})$, $d\sigma/dE_T^{jet1}$, 
$d\sigma/dM_{12}$, $d\sigma/d\bar{\eta}^{jets}$, 
$d\sigma/d|\Delta \eta^{jets}|$ and $d\sigma/dW$ is shown in Figs.\ 4a-h, where
%
\begin{figure}
 \centering
 \includegraphics[width=0.325\columnwidth]{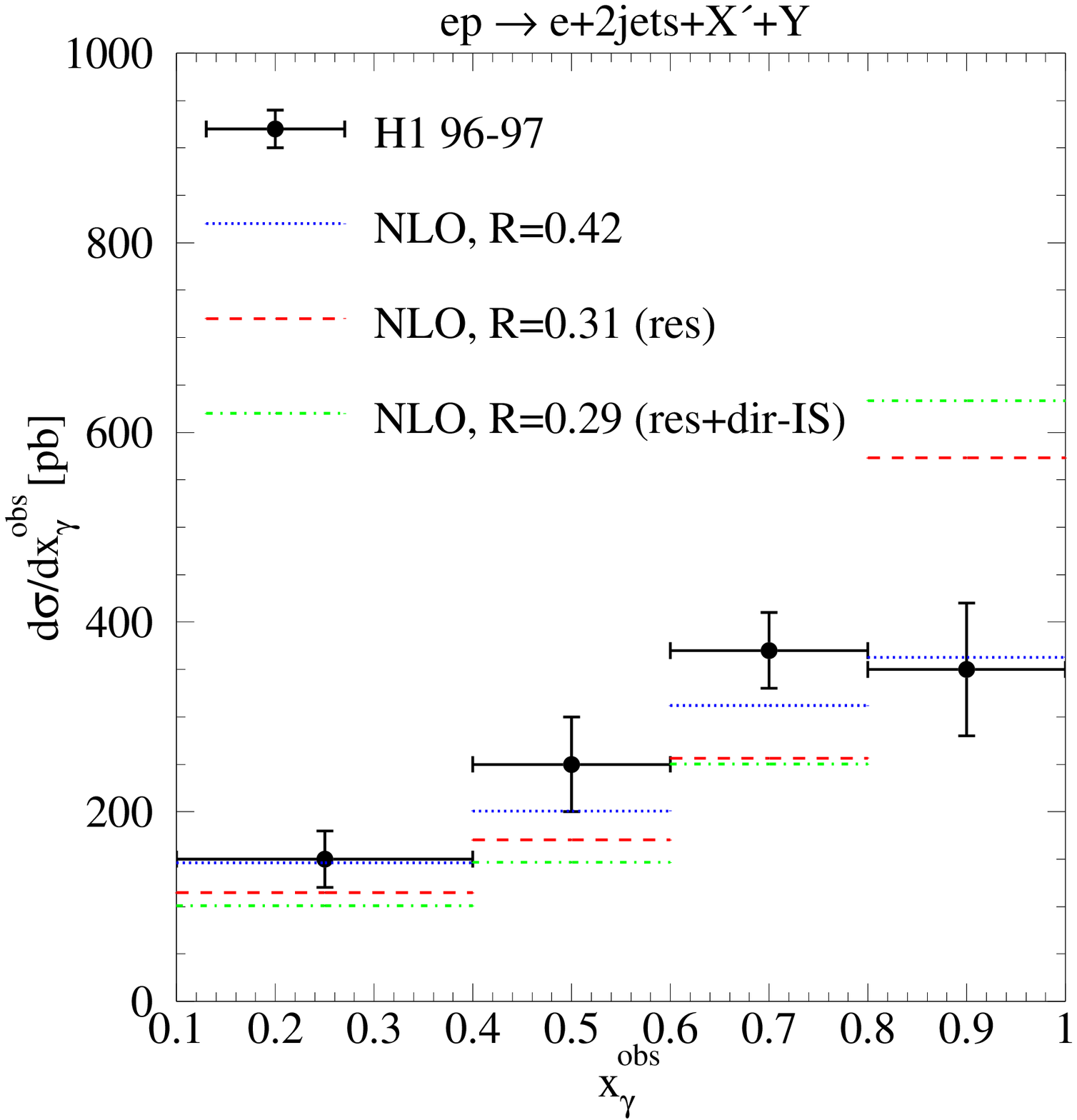}
 \includegraphics[width=0.325\columnwidth]{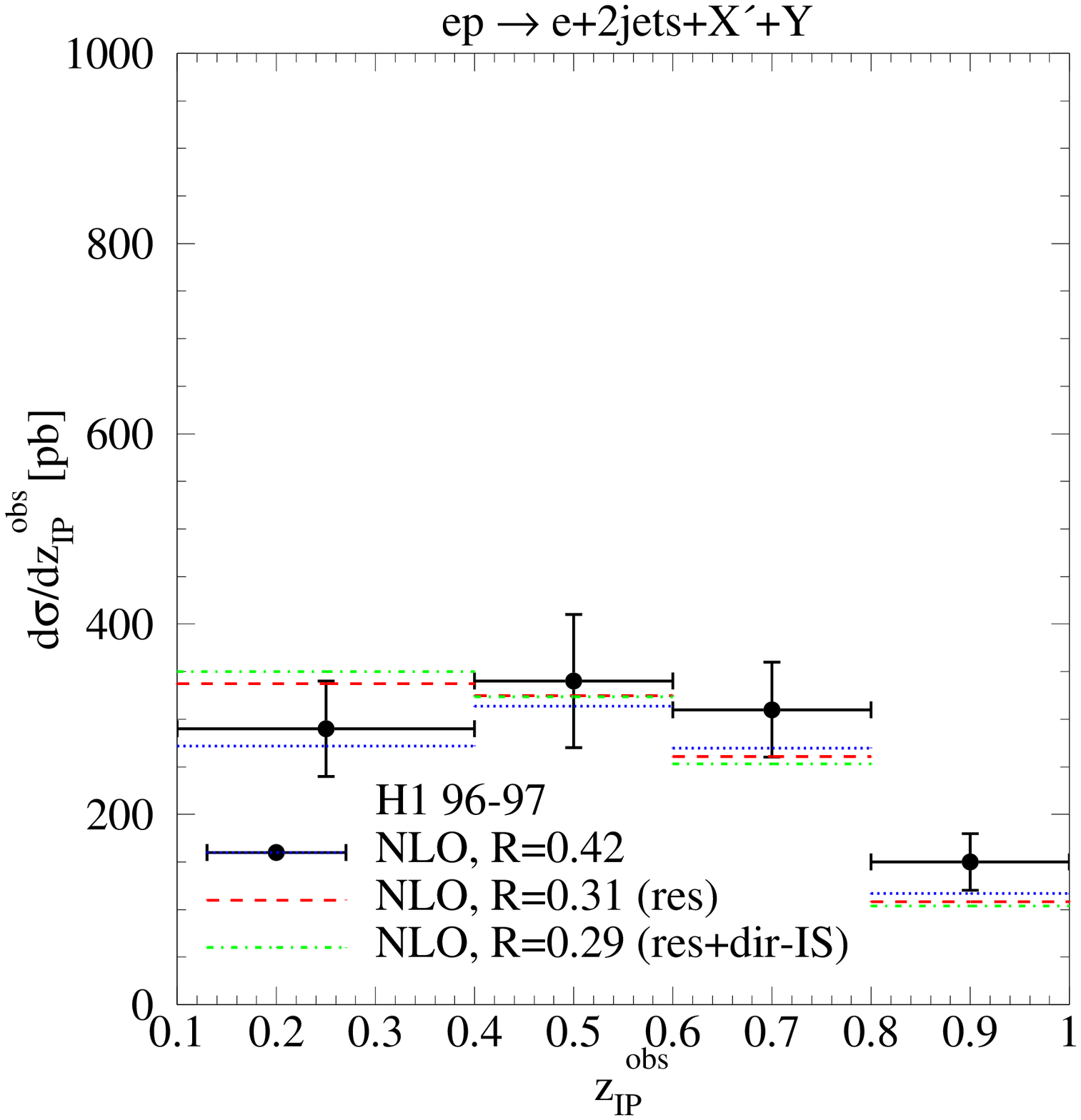}
 \includegraphics[width=0.325\columnwidth]{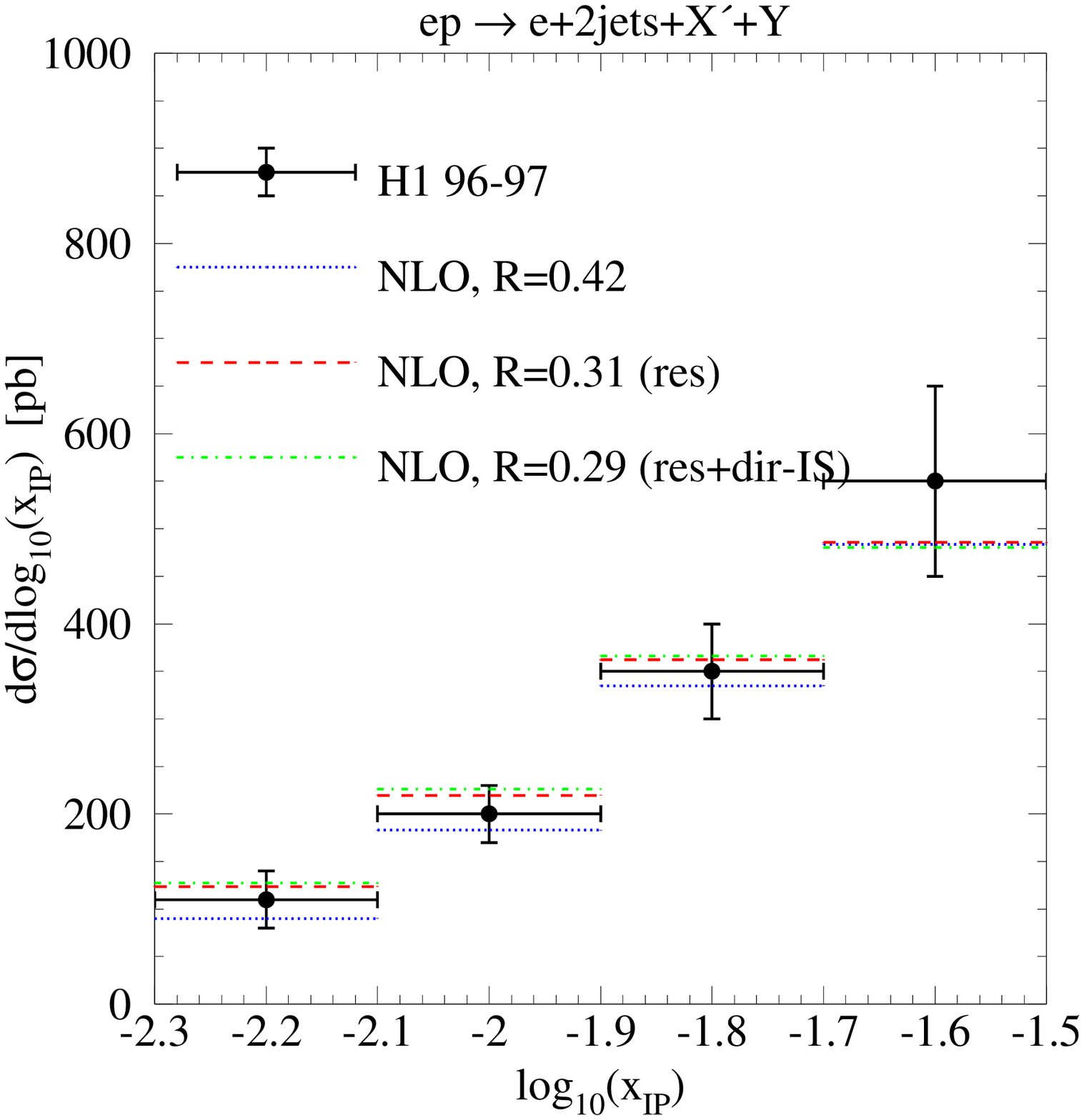}
 \includegraphics[width=0.325\columnwidth]{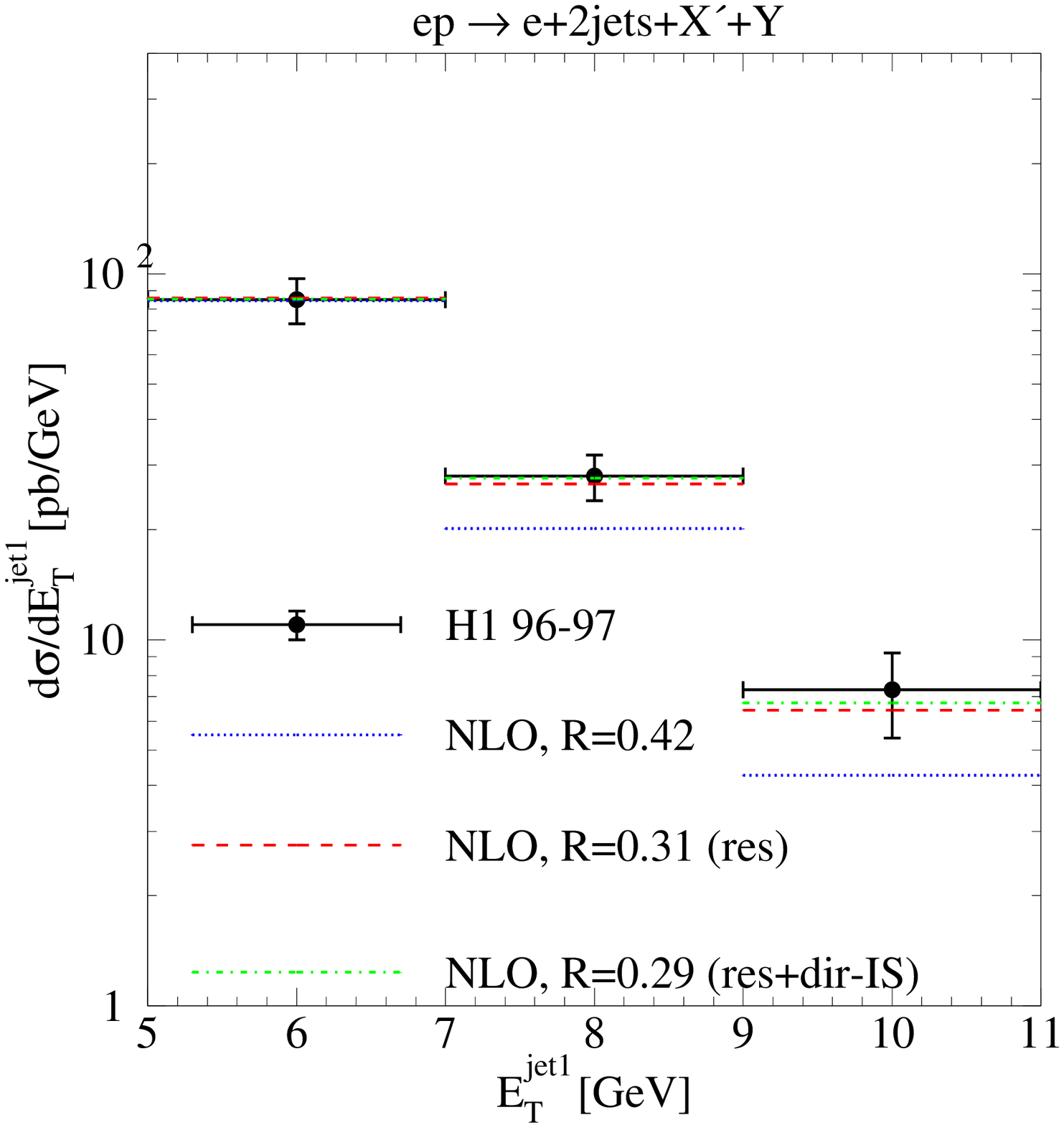}
 \includegraphics[width=0.325\columnwidth]{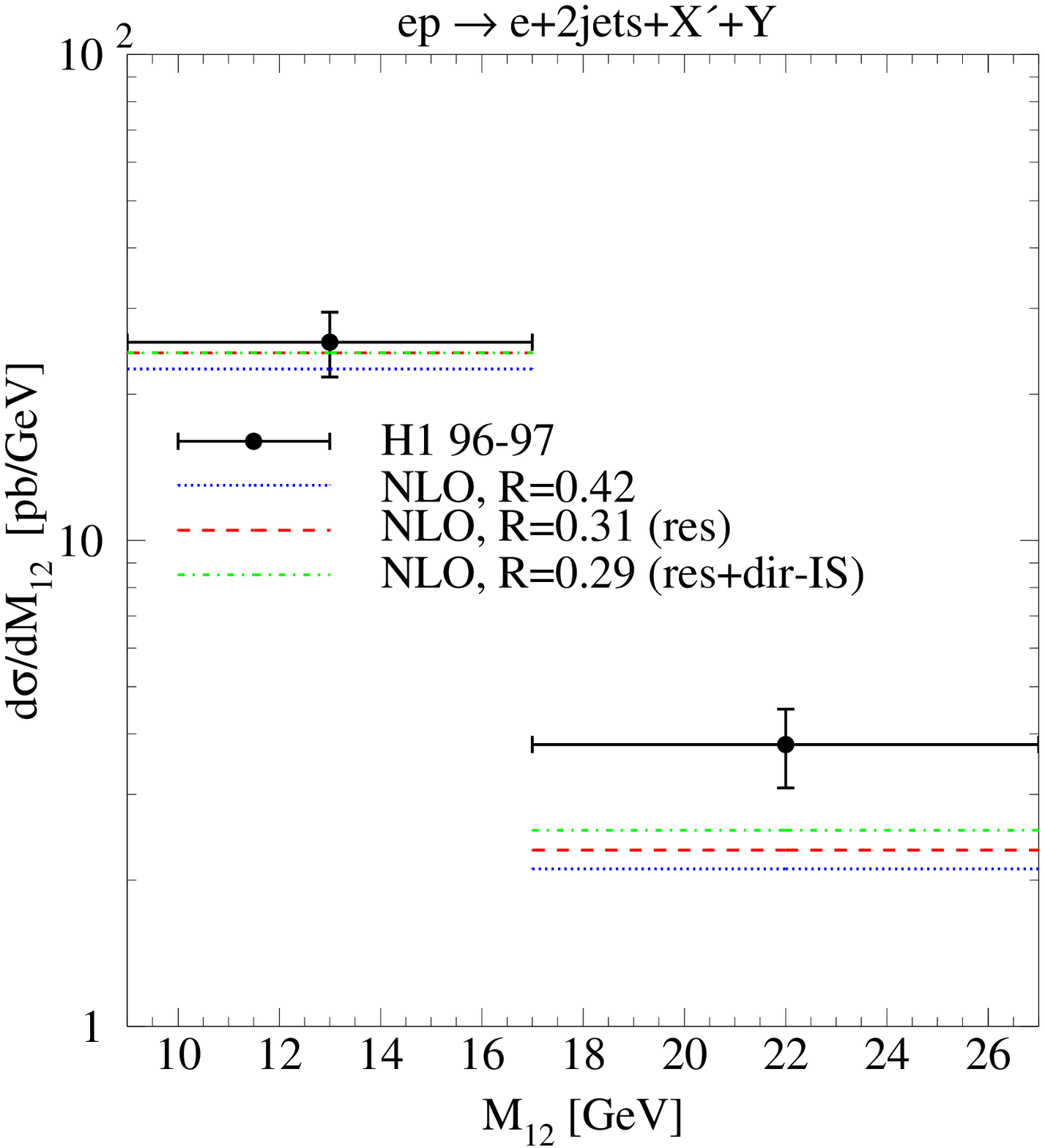}
 \includegraphics[width=0.325\columnwidth]{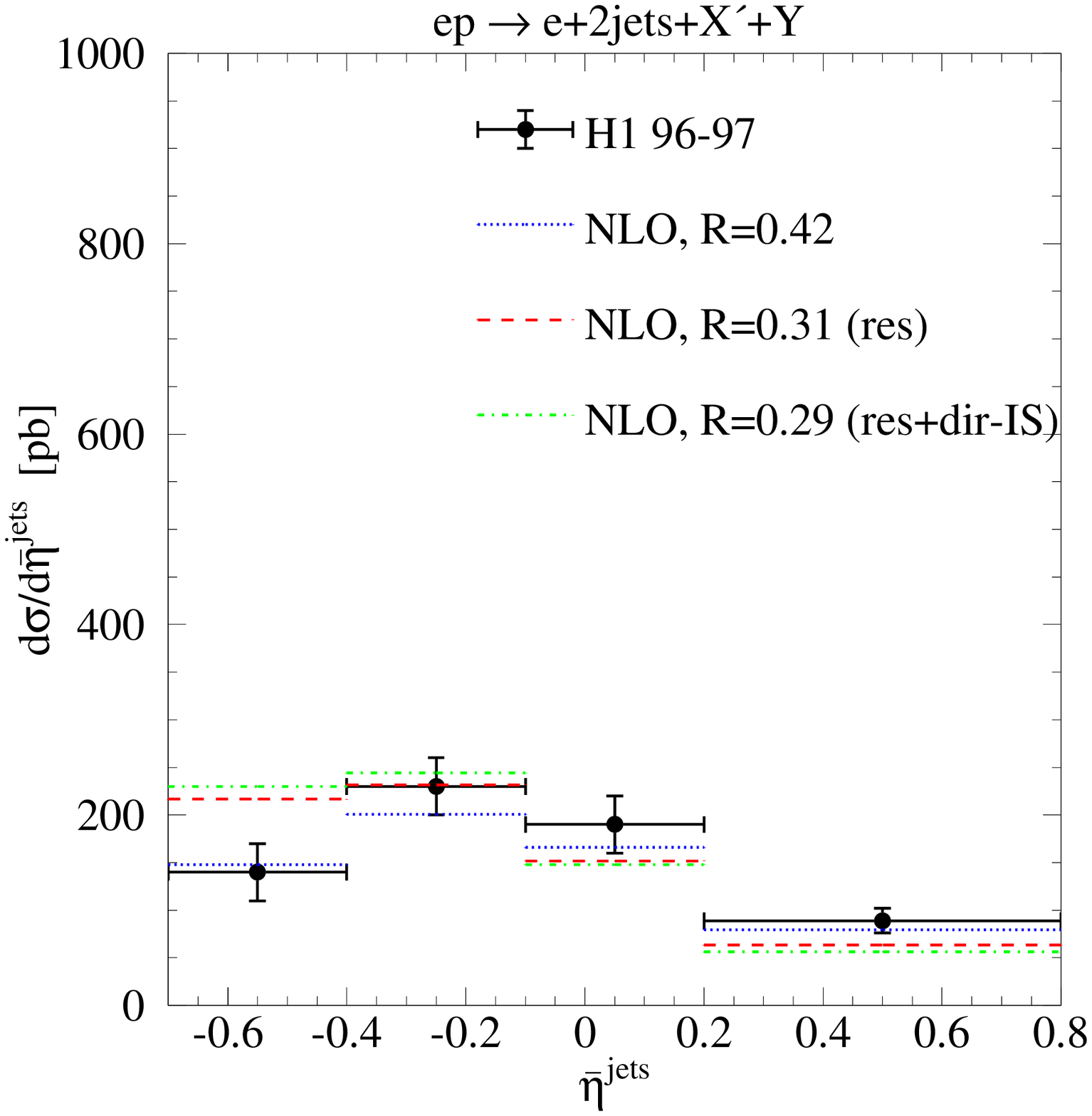}
 \includegraphics[width=0.325\columnwidth]{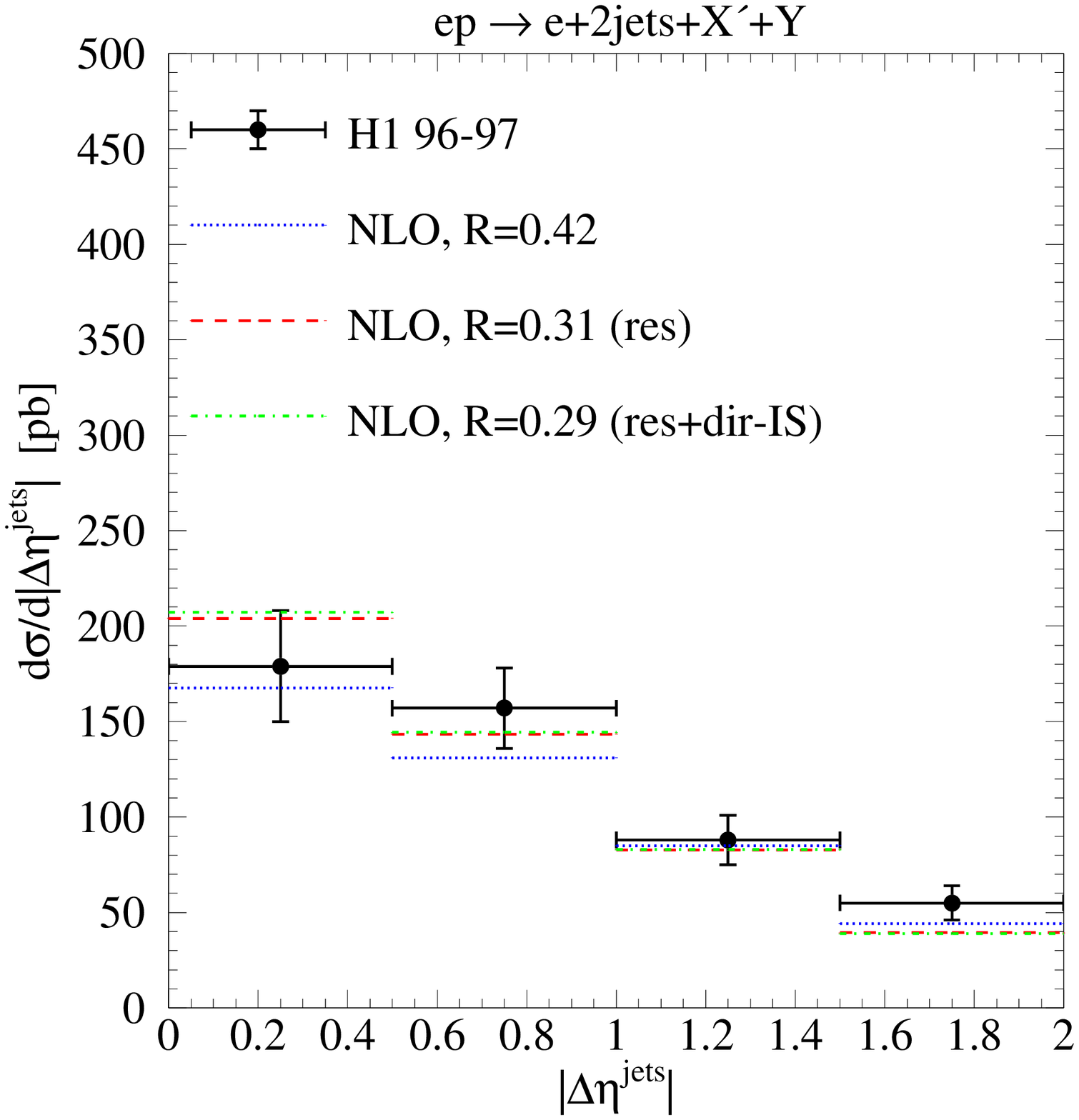}
 \includegraphics[width=0.325\columnwidth]{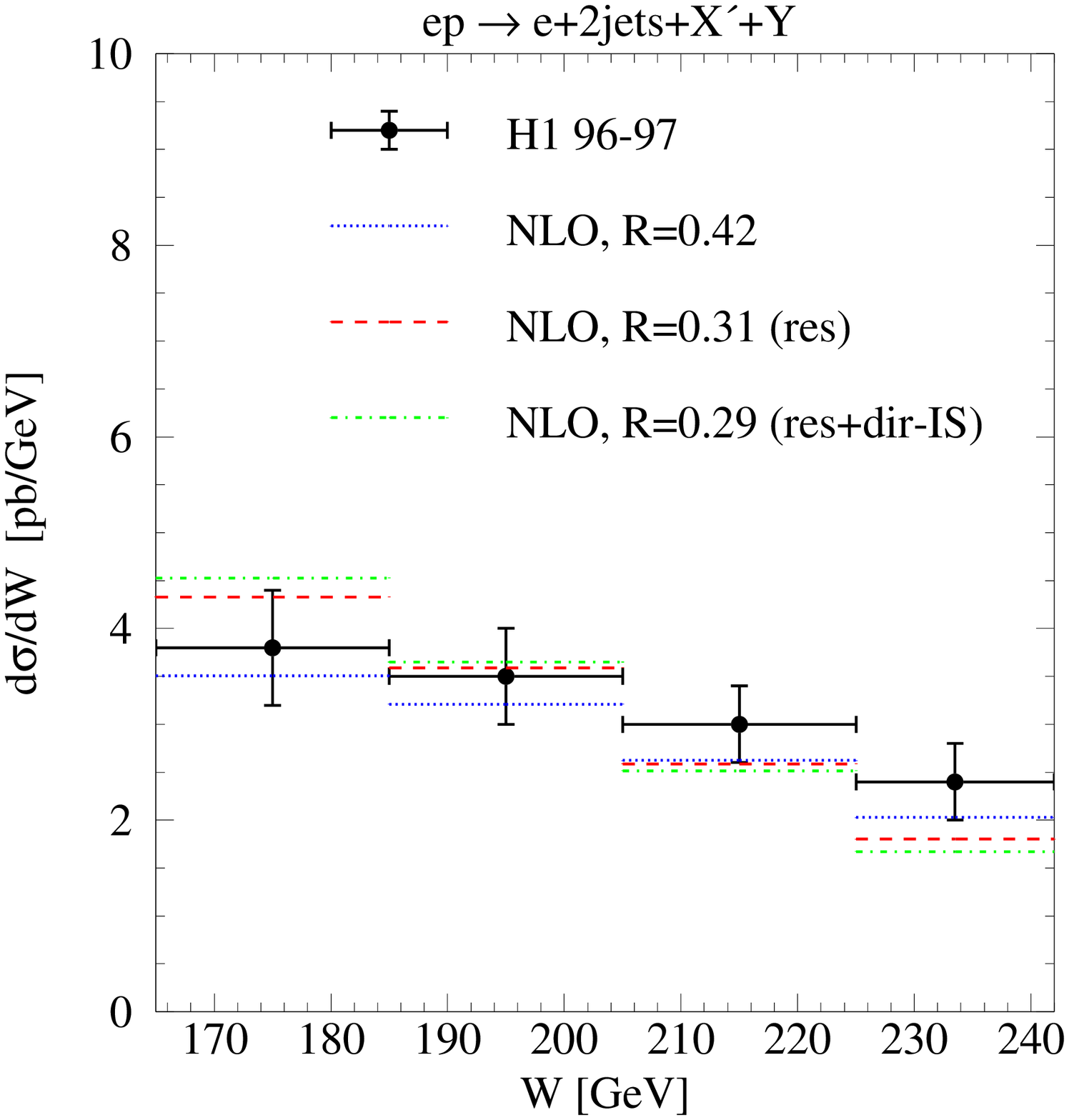}
 \caption{\label{fig:3}Differential cross sections for diffractive dijet
 photoproduction as measured by H1 and compared to
 NLO QCD with global, resolved, and resolved/direct-IS suppression.}
\end{figure}
%
we also have plotted the prediction for the global suppression (direct and
resolved) with $R = 0.42$, already shown in Figs.\ 3a-h.
Looking at Figs.\ 4a-h we can distinguish three groups of results from the 
comparison with the data. In the first group, the cross sections as functions 
of $z_{\p}^{obs}$, $\log_{10}(x_{\p})$, $M_{12}$, $|\Delta \eta^{jets}|$ and 
$W$, the agreement with the global suppression ($R = 0.42$) and the resolved 
suppression ($R = 0.31$ and $R = 0.29$) is comparable. In the second group, 
which consists just of $d\sigma/dE_T^{jet1}$, the agreement is better
for the resolved suppression only. In the third group, 
$d\sigma/dx_{\gamma}^{obs}$ and
$d\sigma/d\bar{\eta}^{jets}$,
the agreement with the resolved suppression is worse than 
with the global suppression. In particular, for $d\sigma/dx_{\gamma}^{obs}$, 
which is usually considered as the characteristic distribution for 
distinguishing global versus resolved suppression, the agreement with 
resolved suppression does not improve. Unfortunately, this cross section has 
the largest hadronic corrections of the order of $(25-30)\%$ \cite{25}. Here, 
the bins with largest $x_{\gamma}^{obs}$ are particularly sensitive to the
hadronic corrections and possible migrations of the data between the two bins.
If we average the cross sections for these two bins, the agreement with the
data point becomes much better. We also  notice, that the predictions for the 
two suppression modes (i) and (ii) are almost the same. The only exception are 
the cross sections for the largest $x_{\gamma}^{obs}$-bin (see Fig.\
4a). In Figs.\ 4a-h the theoretical errors coming from the scale uncertainty
are not shown. If they are taken into account, the difference between experimental
data and theory in Figs.\ 4a and 4f is much less dramatic.
On the
other hand, for the cross section $d\sigma/dE_T^{jet1}$ the agreement improves 
considerably with the suppression of the resolved part only (note the 
logarithmic scale in Fig.\ 4d). Here, of course, we must admit that the 
suppression factor could be $E_T$-dependent, although we do not know of any
mechanism, which could cause such a $E_T^{jet}$-dependence of the suppression.
We remark that this $E_T^{jet}$-dependence of the global suppression is also
visible in the H1 analysis of \cite{25}.
 
We also checked for two distributions whether the predictions for resolved 
suppression depend on the chosen diffractive PDFs. For this purpose we have 
calculated for the two cases $d\sigma/dz_{\p}^{obs}$ and $d\sigma/dE_T^{jet1}$ 
the cross sections with the `H1 2006 fit A' parton distributions
\cite{4}. The results are compared in Figs.\ 5a and b to the results with
%
\begin{figure}
 \centering
 \includegraphics[width=0.325\columnwidth]{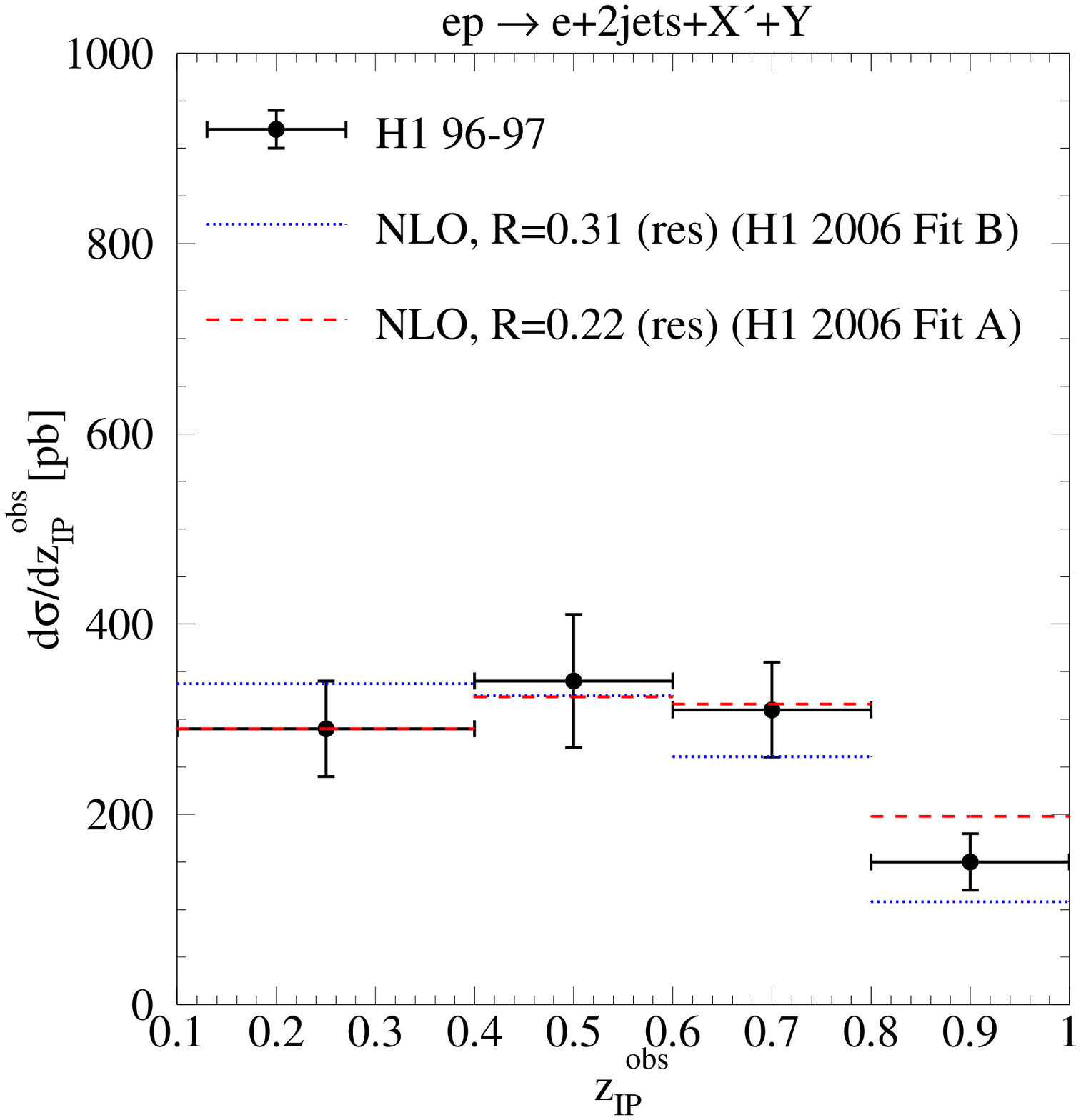}
 \includegraphics[width=0.325\columnwidth]{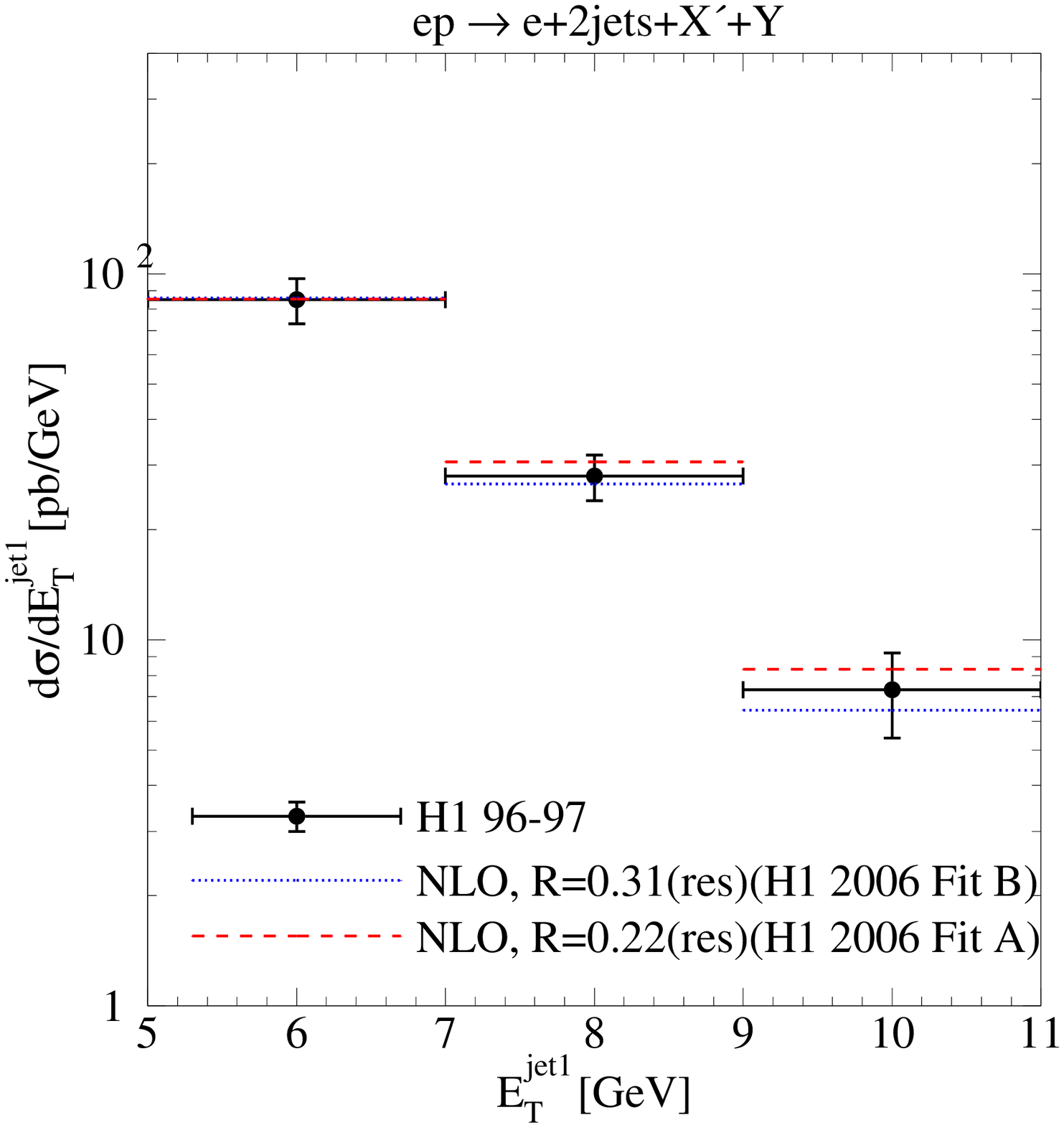}
 \caption{\label{fig:4}Differential cross sections for diffractive dijet
 photoproduction as measured by H1 and compared to
 NLO QCD with resolved suppression and two different DPDFs.}
\end{figure}
%
the `H1 2006 fit B' and the experimental data. Of course, since the `H1 2006
fit A'
PDFs have a larger gluon component at large $\beta$, the cross sections are
larger and therefore need a larger suppression factor $R = 0.22$. From Figs.\
5a and b we conclude that there is no appreciable dependence on the chosen
DPDFs. Note that in Fig.\ 5b the cross section for the smallest
$E_T^{jet1}$-bin
has been fitted to determine the suppression factor. In total, we are tempted to conclude from the comparisons in  Figs.\ 4a-h that the predictions with a 
resolved-only (or resolved+direct-IS) suppression are consistent with the H1 
data \cite{25}. The only exceptions are two bins in the $x_{\gamma}^{obs}$ and 
one bin in the $\bar{\eta}^{jets}$-distribution.

\subsection{Comparison with ZEUS data}

In this subsection we shall compare our predictions with the final analysis
of the ZEUS data, which was published just recently \cite{27}. The
kinematic cuts are given in Tab.\ 2. There are the following differences to 
the H1 cuts in Tab.\ 1: First the upper cut on $Q^2$ is larger. Second there is
a larger range in the variable $y$ and the upper cut on $x_{\p}$ is slightly
smaller. The most important change is the larger $E_T^{jet1(2)}$ cut, namely
$E_T^{jet1(2)} >$ 7.5 (6.5)
GeV, which leads to smaller cross sections. Also the
cut on $|t|$ is different.
The different cuts
on $Q^2$ and $|t|$  have little influence. For example, the larger $|t|$-cut 
in Tab.\ 2 as compared to Tab.\ 1 increases the cross section only by $0.2\%$.
The constraint on $M_Y$ is not explicitly given in the ZEUS publication 
\cite{27}. They give the cross section for the case that the diffractive final 
$Y$ state consists only of the proton. For this they correct their measured 
cross section by subtracting in all bins the estimated contribution of a
proton-dissociative background of $16\%$. When comparing to the theoretical
predictions with the DPDFs from the H1 2006 fits, they multiply the theoretical
cross section with a (slightly different) factor of $0.87$ in order to correct
for the
proton-dissociative contributions, which are contained in these DPDFs by
requiring $M_Y < 1.6$ GeV. We do not follow this procedure. Instead we leave 
the theoretical cross sections unchanged, i.e.\ they contain a
proton-dissociative contribution with $M_Y < 1.6$ GeV and multiply the ZEUS 
cross sections by $1.15$ to include the proton-dissociative contribution. 
In this comparison we shall follow the same strategy as before. Before we do 
this, we correct our theoretical prediction by the hadronization corrections
reported in \cite{27}. We first compare to the 
predictions with no suppression ($R = 1$) and then determine a suppression
factor by fitting $d\sigma/dE_T^{jet1}$ at the smallest $E_T^{jet1}$-bin.
Then we compare to the cross sections as a function of the seven observables 
$x_{\gamma}^{obs}$, $z_{\p}^{obs}$, $x_{\p}$, $E_T^{jet1}$, $y$, $M_X$ and
$\eta^{jet1}$
instead of the eight variables in the H1 analysis. The distribution in $y$ is 
equivalent to the $W$-distribution in \cite{25}. The ZEUS collaboration
have also published experimental measurements in the two regions
$x_\gamma<(\geq)~0.75$, which do however not consider here due to space
limitations.

The theoretical predictions
for these differential cross sections with no suppression factor ($R = 1$) are
shown in Figs.\ 6a-g, together with their scale errors and compared to
%
\begin{figure}
 \centering
 \includegraphics[width=0.325\columnwidth]{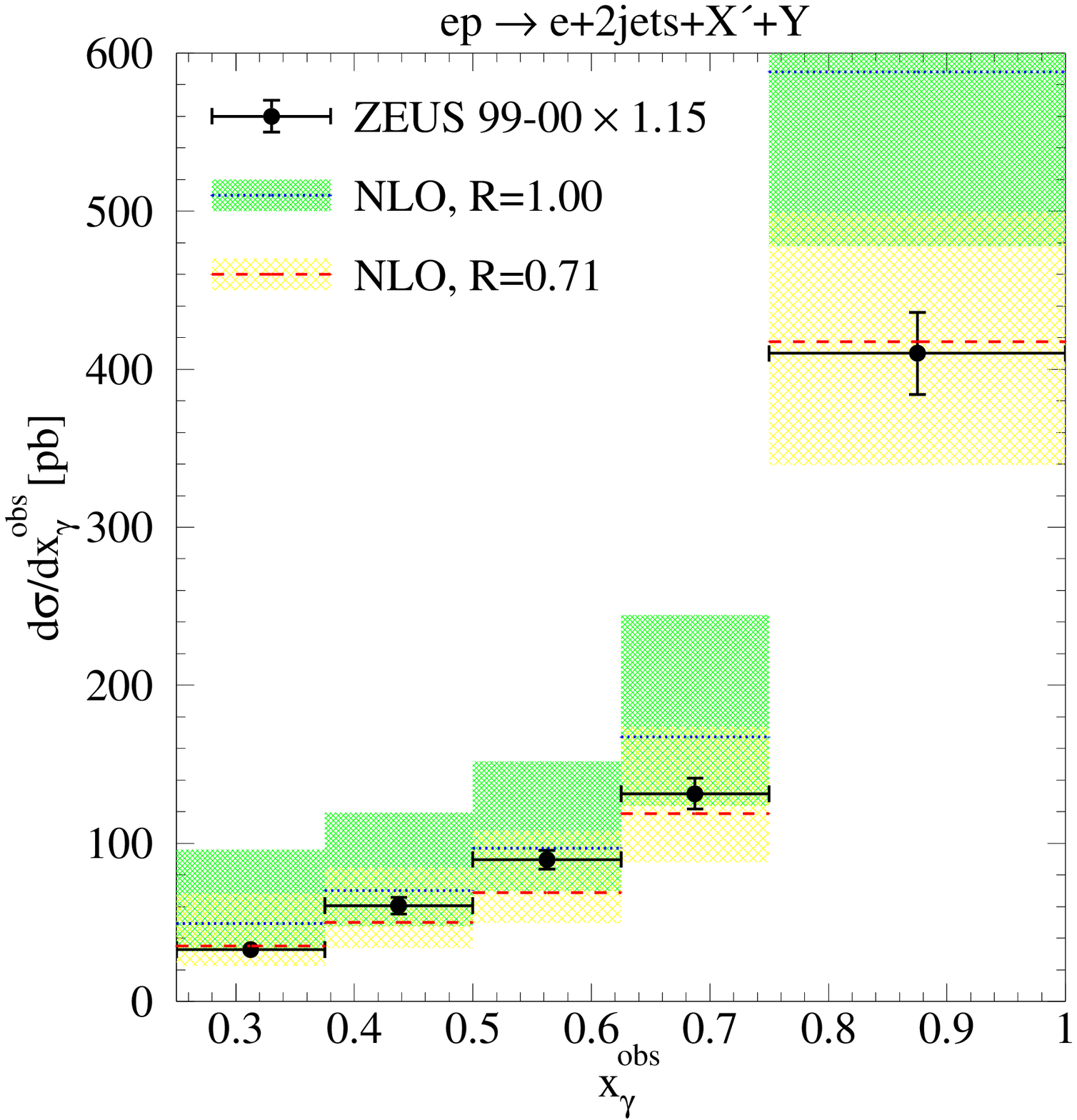}
 \includegraphics[width=0.325\columnwidth]{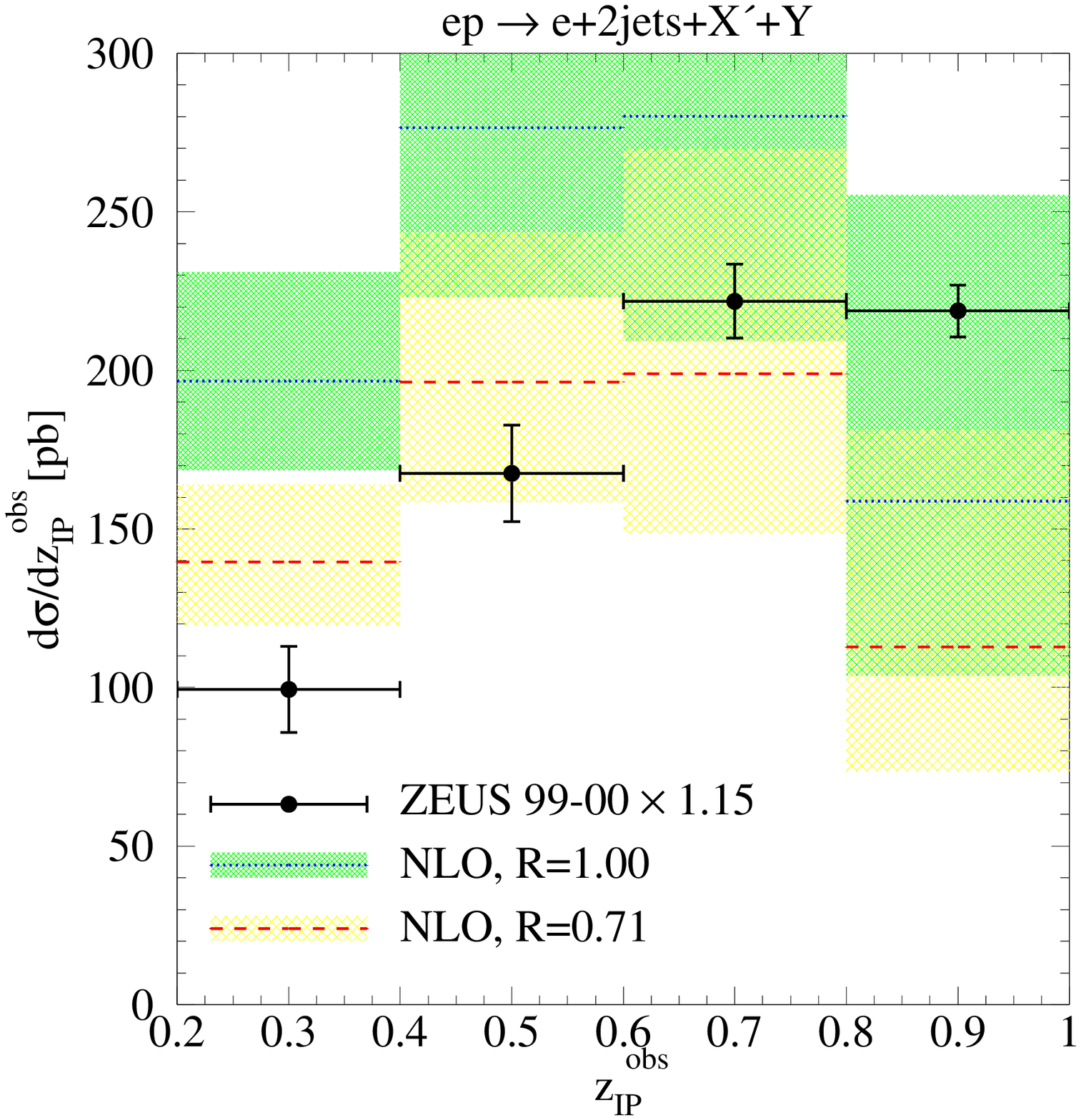}
 \includegraphics[width=0.325\columnwidth]{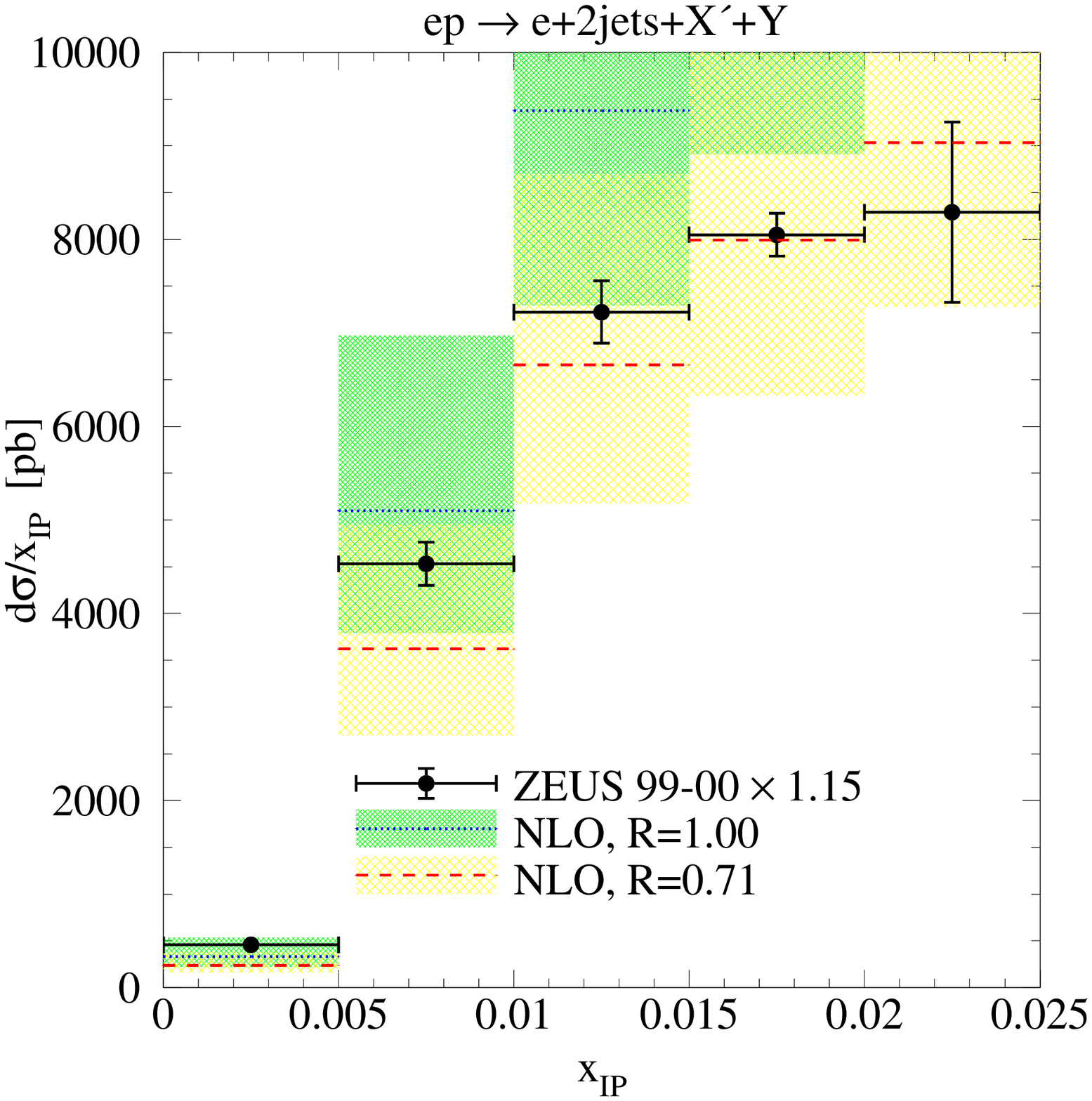}
 \includegraphics[width=0.325\columnwidth]{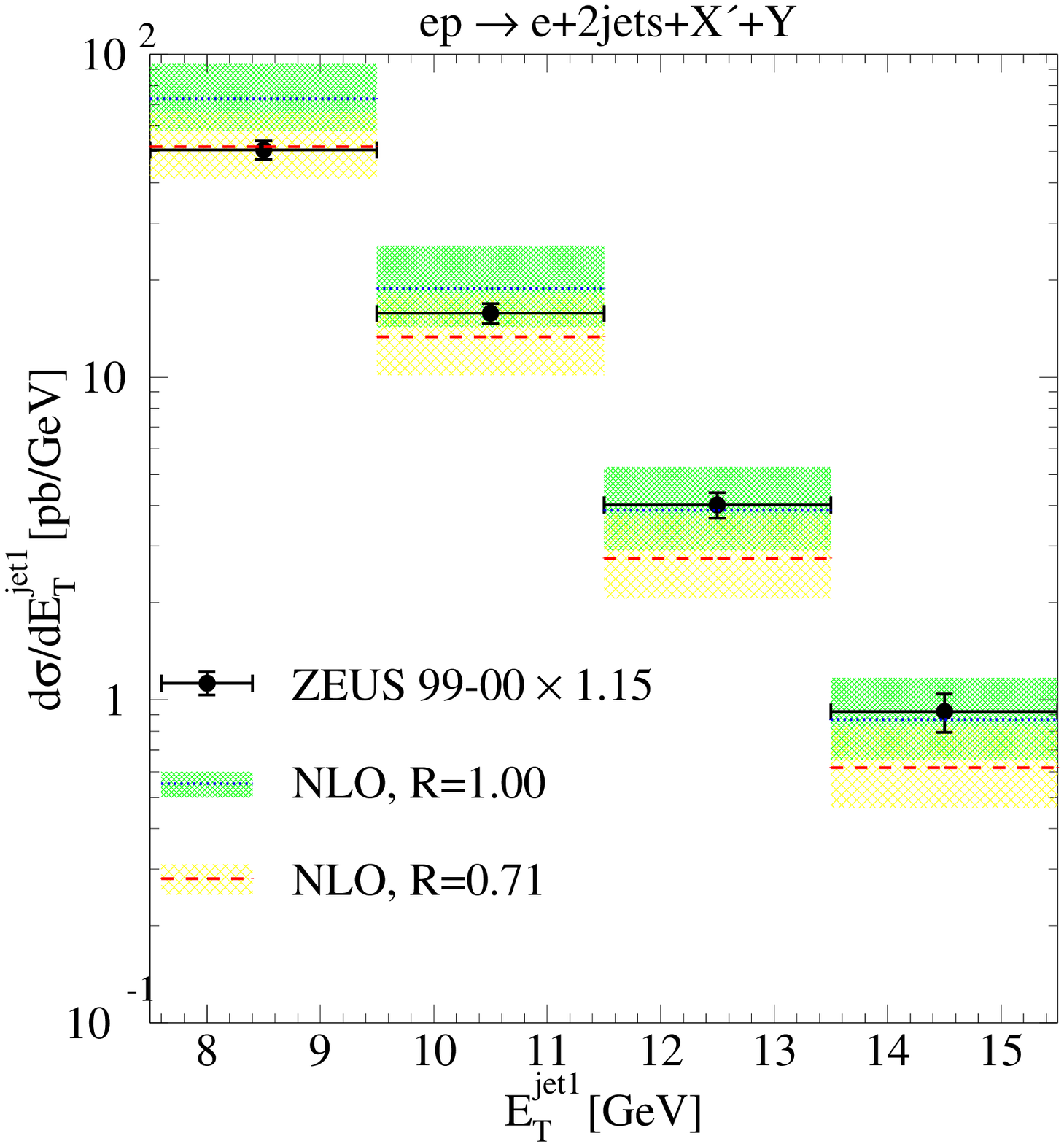}
 \includegraphics[width=0.325\columnwidth]{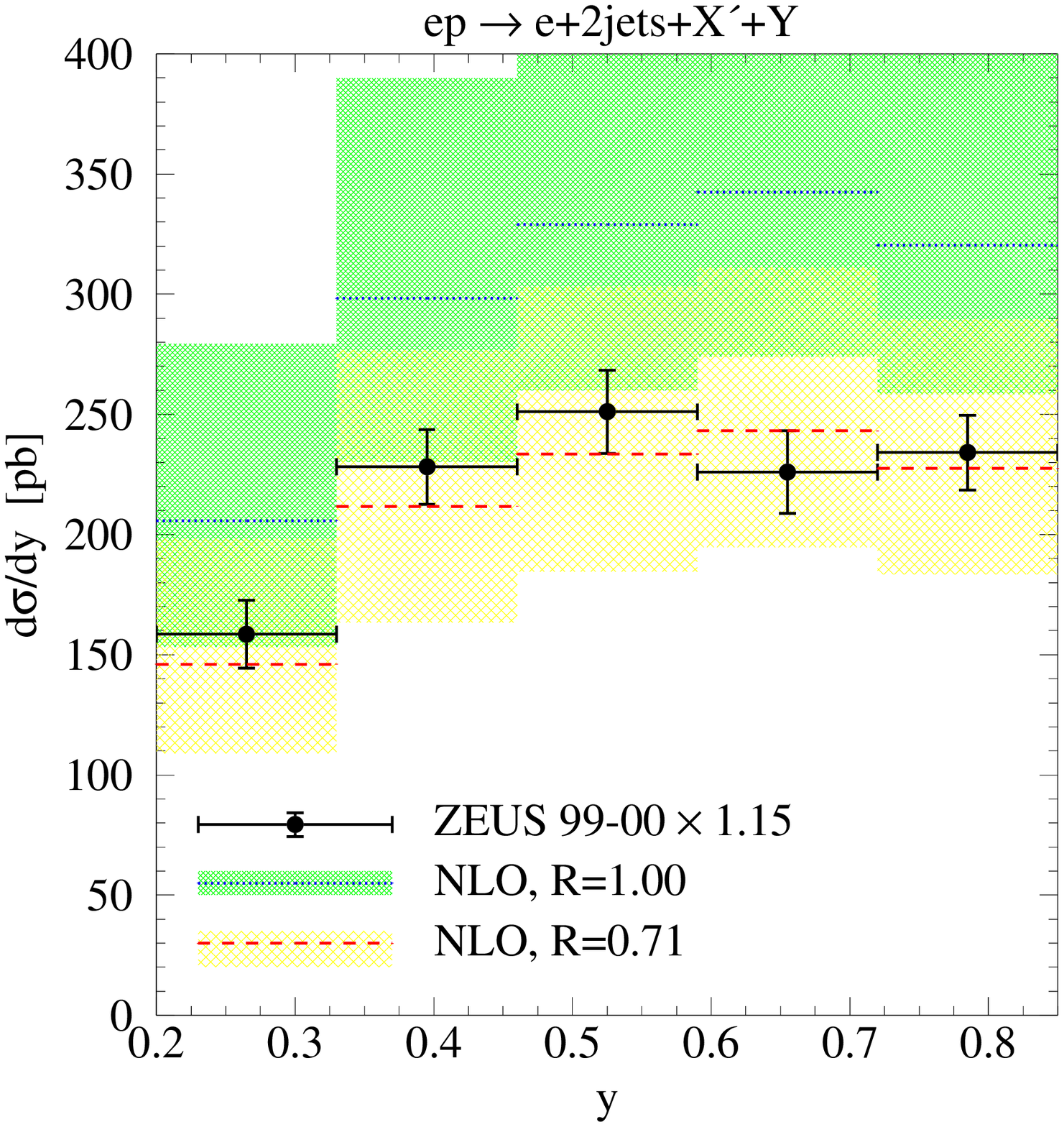}
 \includegraphics[width=0.325\columnwidth]{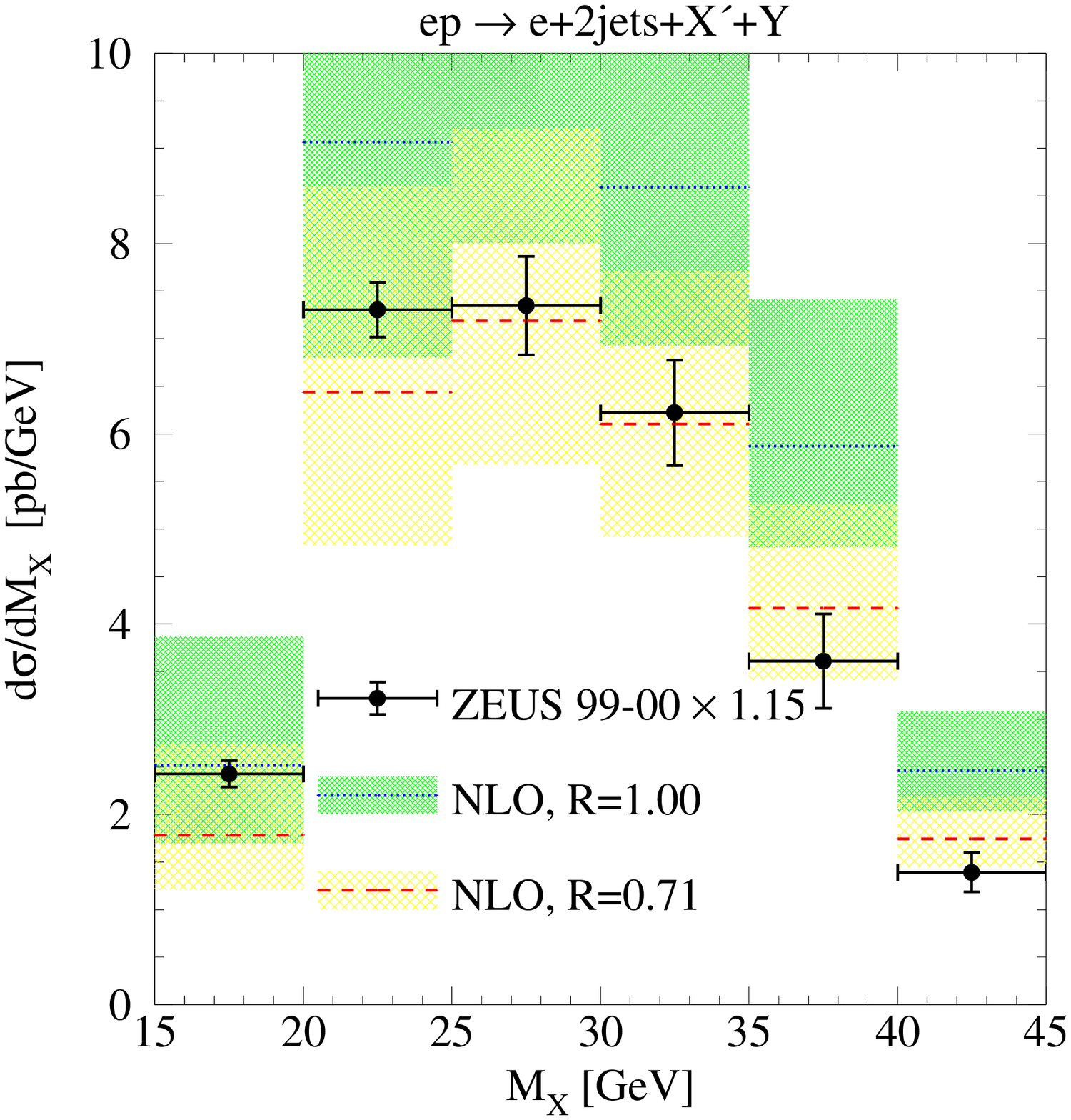}
 \includegraphics[width=0.325\columnwidth]{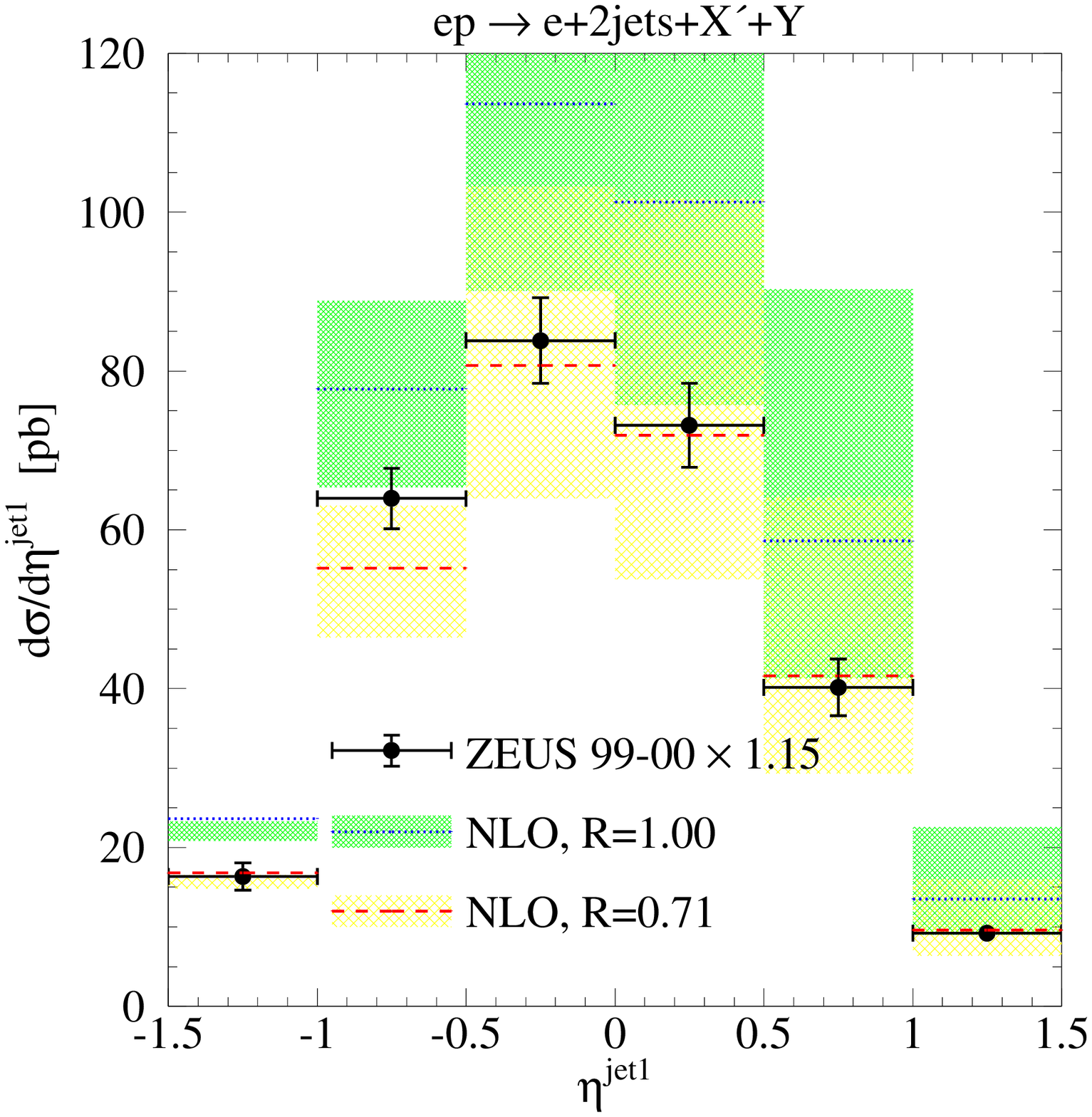}
 \caption{\label{fig:8}Differential cross sections for diffractive dijet
 photoproduction as measured by ZEUS and compared to
 NLO QCD without ($R=1$) and with ($R=0.71$) global suppression (color online).}
\end{figure}
%
the ZEUS data points. Except for the $x_{\gamma}^{obs}$ and $E_T^{jet1}$
distributions, most of the data points lie outside the
theoretical error bands for $R = 1$. In particular, in Figs.\ 6b, c, e, f and
g, 2, 3, 4, 4 and 5 points lie outside. This means that most of the data
points disagree with the unsuppressed prediction. Next, we determine the
suppression factor from the measured $d\sigma/dE_T^{jet1}$ at the lowest 
$E_T^{jet1}$-bin, 7.5 GeV $<E_T^{jet1}<9.5$ GeV, and obtain 
$R = 0.71 \pm 0.06$.
The indicated error corresponds again to the experimental uncertainty, while we
show in the figures explicitly the theoretical uncertainty. 
This means that the suppression factor from the ZEUS data 
is larger than the one obtained from the analysis of the H1 data, which is 
actually consistent with the $d\sigma/dE_T^{jet1}$ for the second bin in 
Fig.\ 3b. Here the cross section is approximately larger by a factor of $1.8$
than the prediction with $R=0.41$.
If we now check how the predictions for $R = 0.71$ compare to the data
points inside the theoretical errors, we observe from Figs.\ 6a-g
that with the exception of $d\sigma/dz_{\p}^{obs}$ (largest bin), the data 
points agree with the predictions inside the theoretical error band. This is 
quite consistent with the H1 analysis, discussed in the previous subsection, 
and leads to the conclusion that also the ZEUS data agree much better with the 
suppressed predictions than with the unsuppressed one. In particular, the 
global suppression factor approximately agrees with the global suppression 
factor, which one would expect from the analysis of the H1 data at the second 
smallest $E_T^{jet1}$-bin.

Similarly as in the previous section we compared the ZEUS data also with the
assumption that the suppression results only from the resolved cross
section. Here we consider again the two versions: (i) only resolved
suppression (res) and (ii) resolved plus direct suppression of the
initial-state
singular part (res+dir-IS). For these two models we obtain the
suppression factors $R = 0.53$ and $R = 0.45$, respectively, where these
suppression
factors are again obtained by fitting the data point at the first bin of
$d\sigma/dE_T^{jet1}$. The comparison to the global suppression with $R=0.71$ 
and to the data is shown in Figs.\ 7a-g.
%
\begin{figure}
 \centering
 \includegraphics[width=0.325\columnwidth]{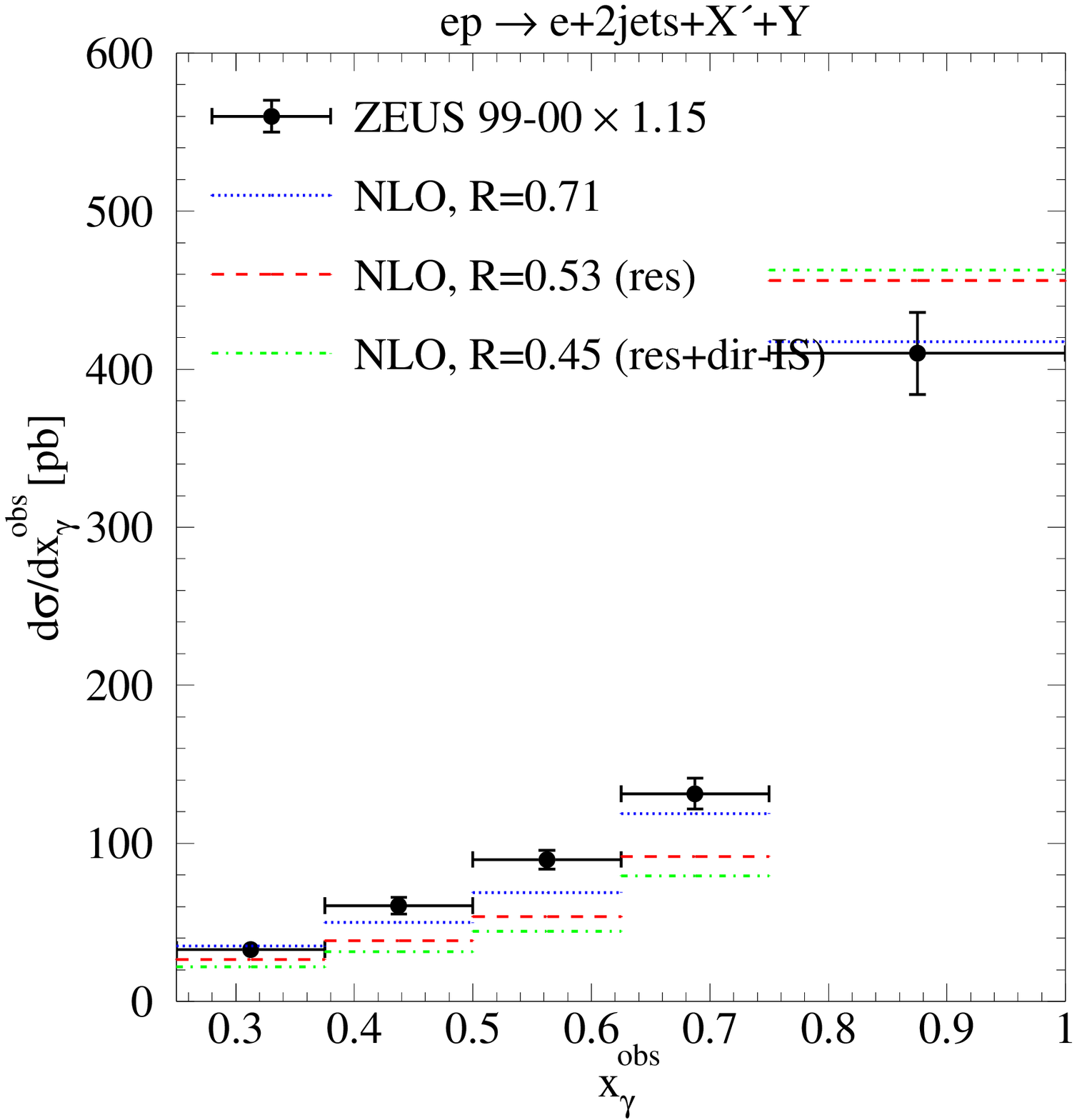}
 \includegraphics[width=0.325\columnwidth]{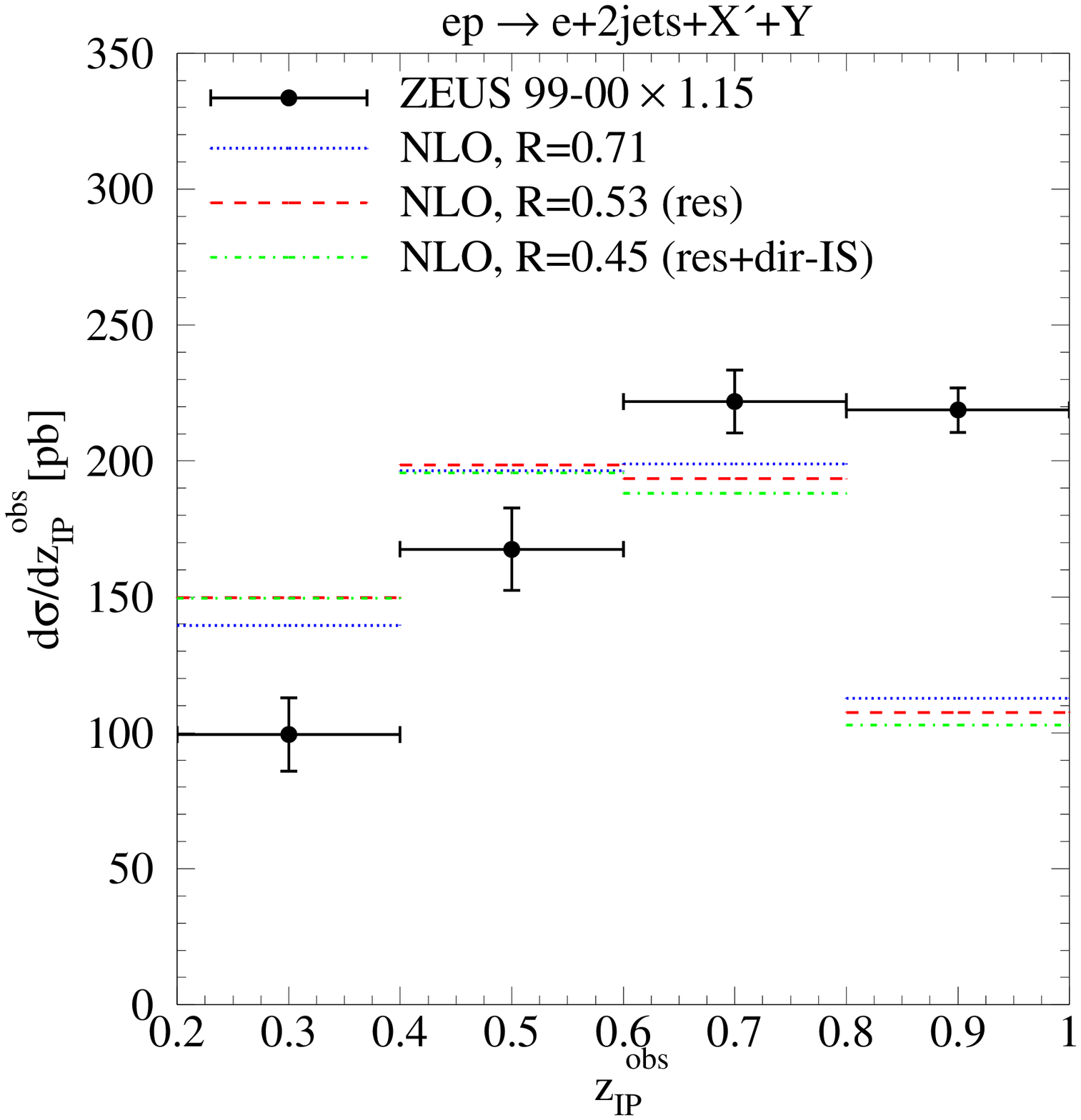}
 \includegraphics[width=0.325\columnwidth]{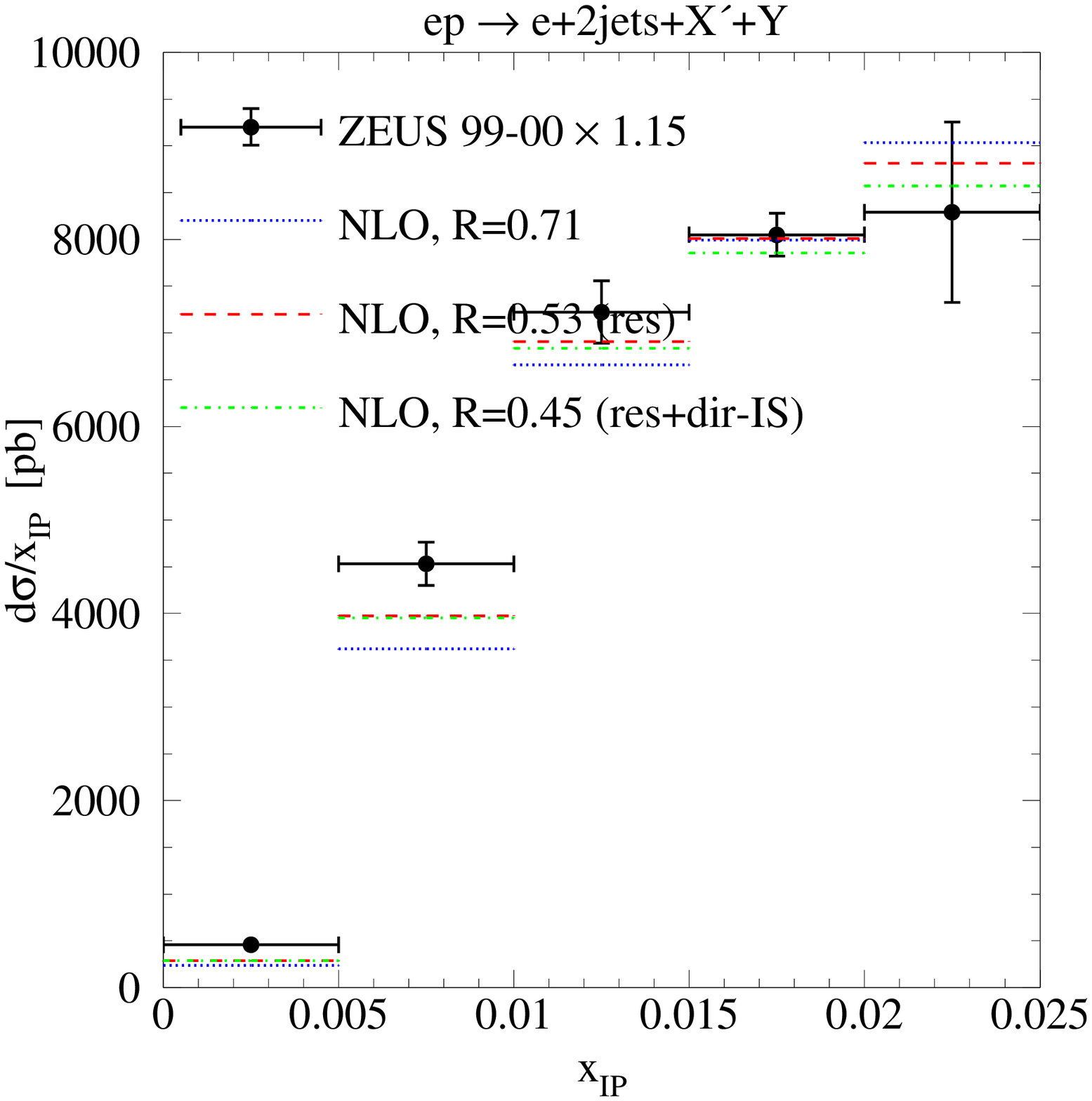}
 \includegraphics[width=0.325\columnwidth]{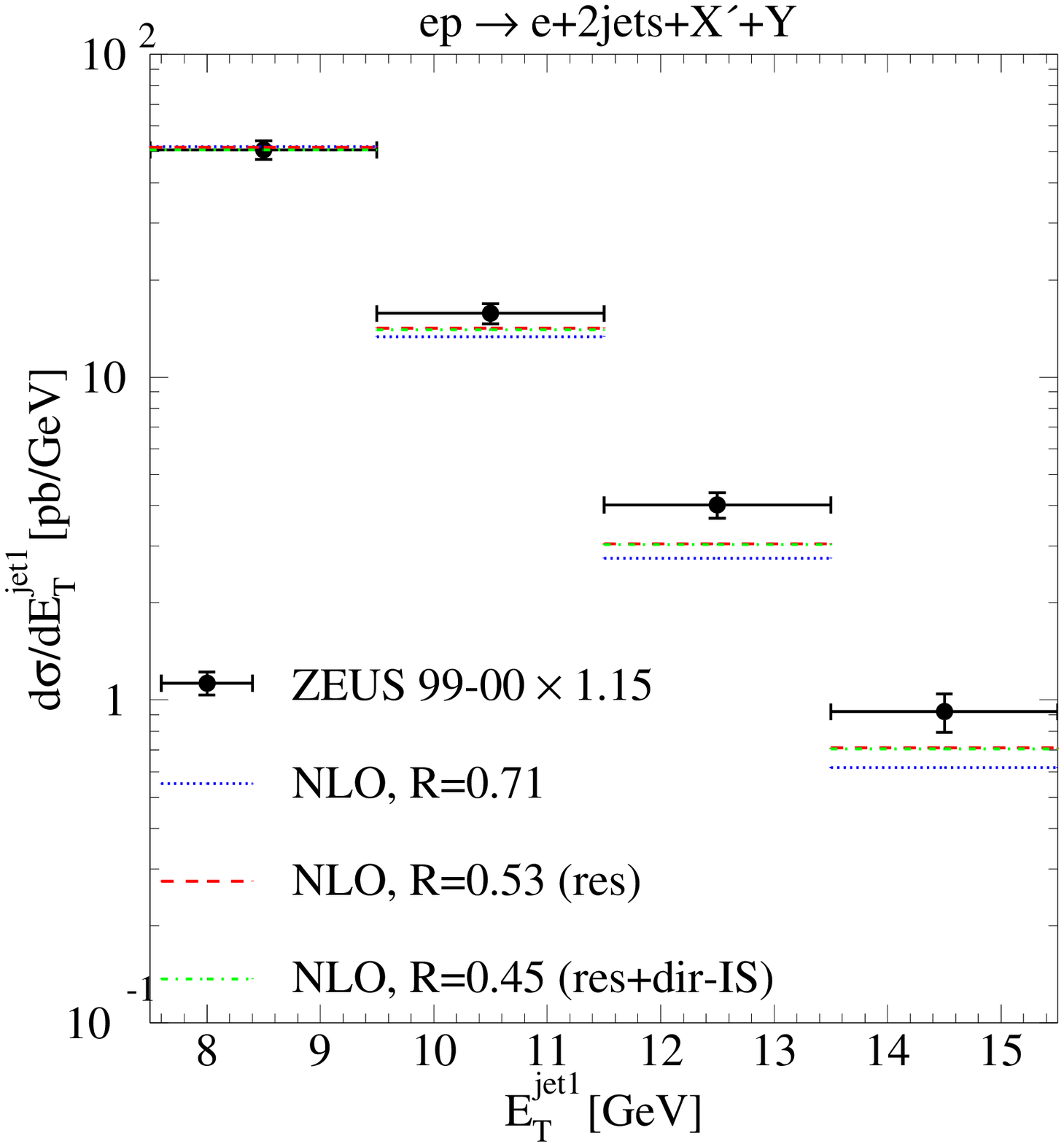}
 \includegraphics[width=0.325\columnwidth]{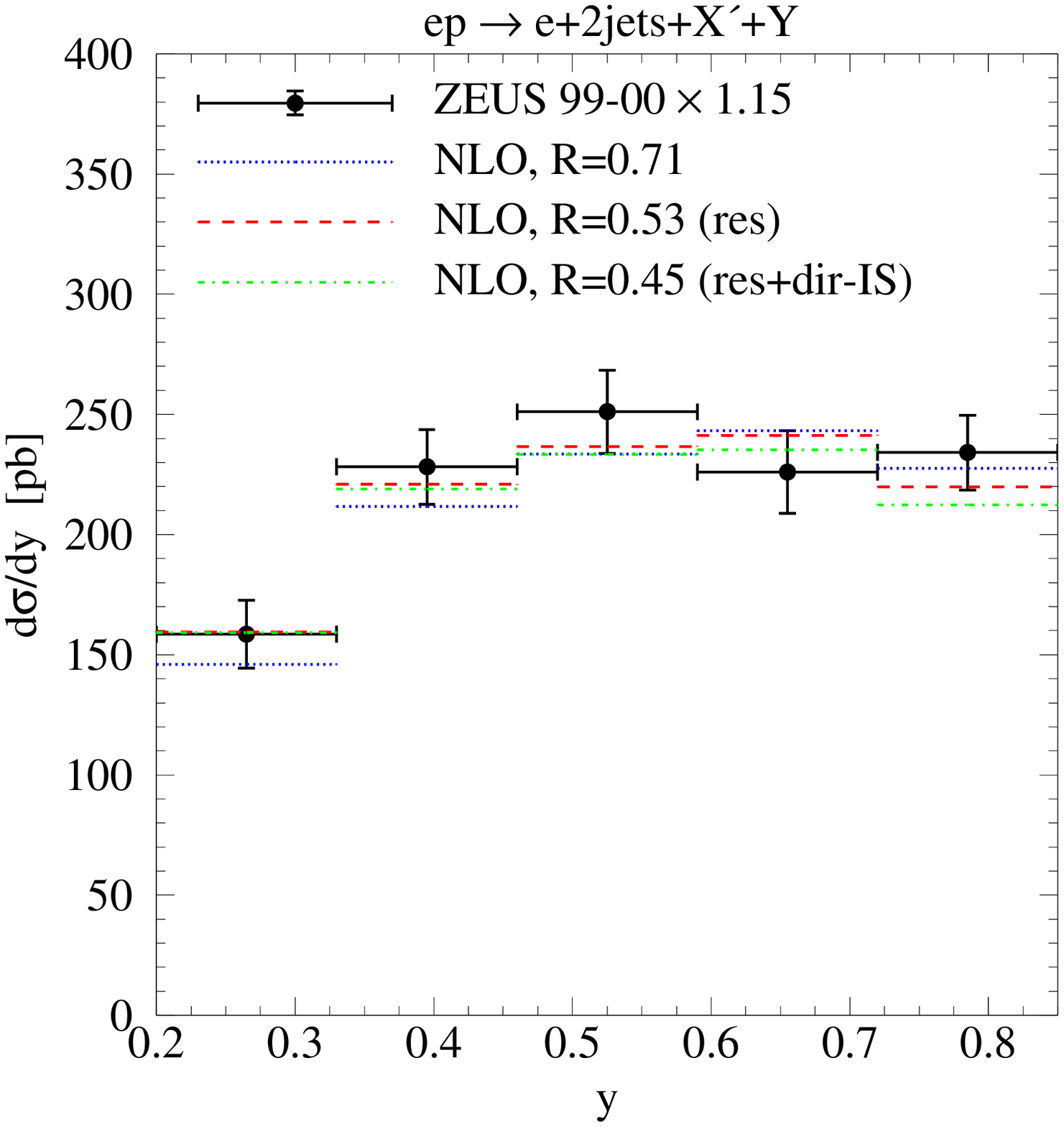}
 \includegraphics[width=0.325\columnwidth]{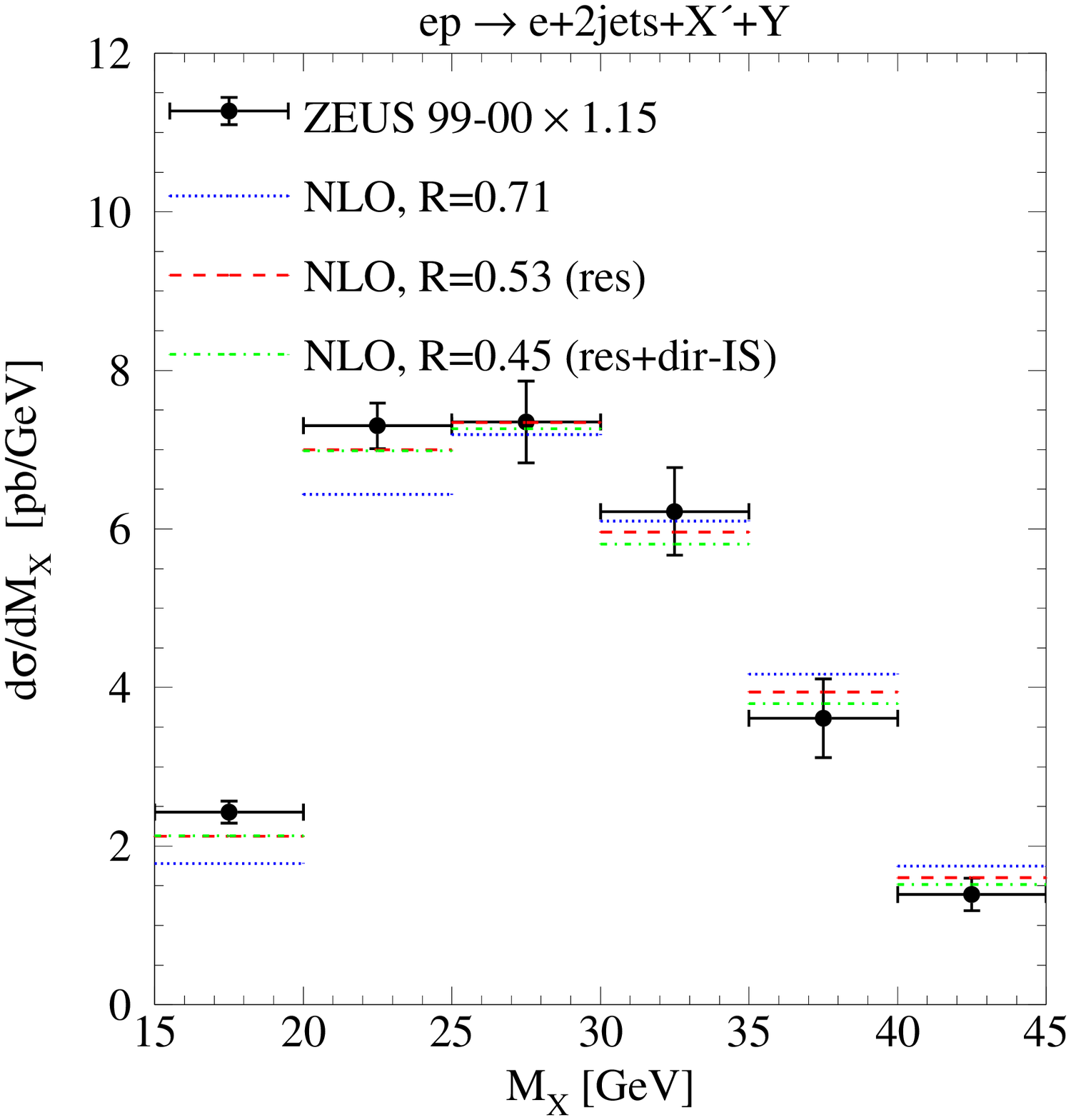}
 \includegraphics[width=0.325\columnwidth]{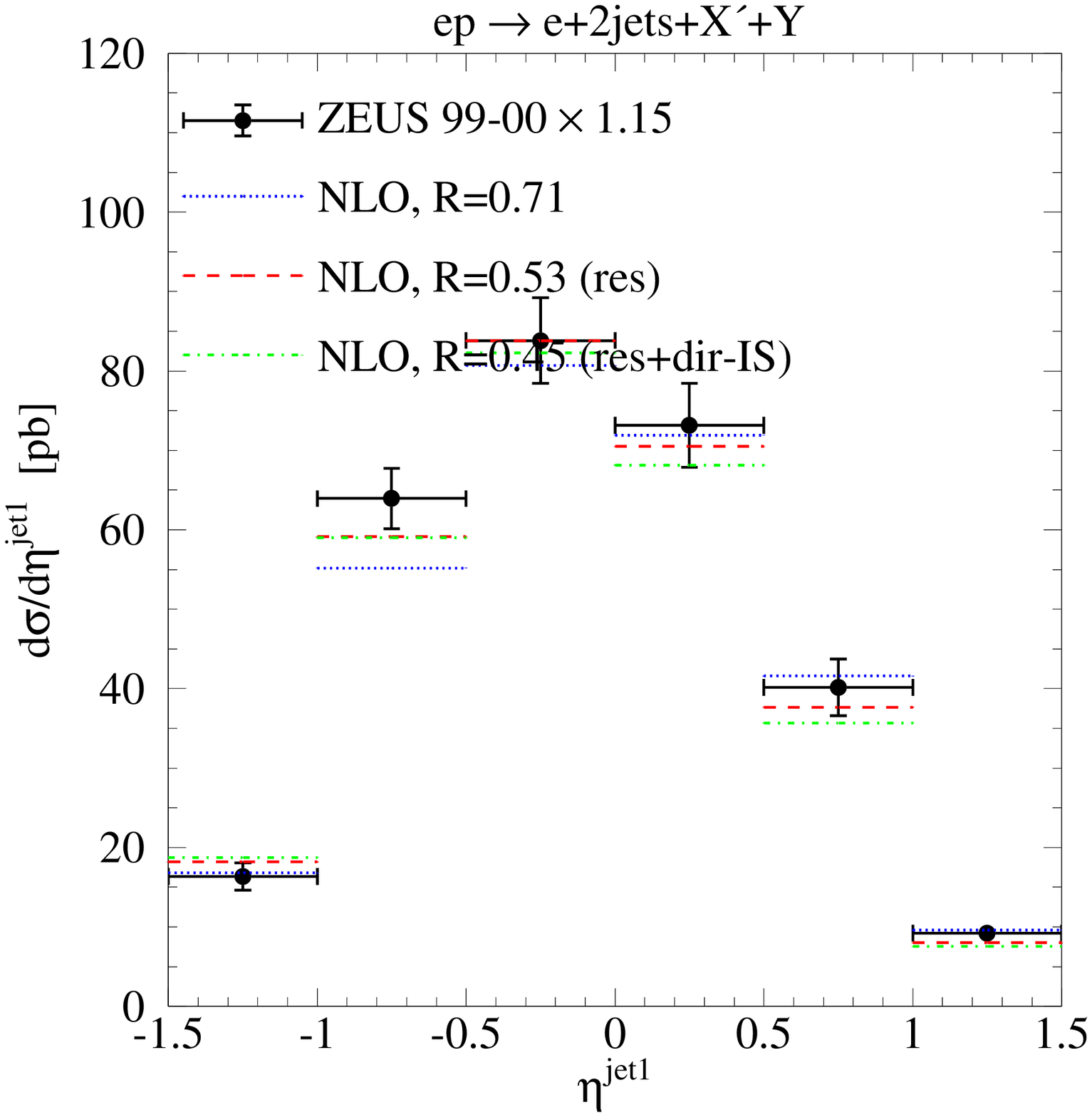}
 \caption{\label{fig:9}Differential cross sections for diffractive dijet
 photoproduction as measured by ZEUS and compared to
 NLO QCD with global, resolved, and resolved/direct-IS suppression.}
\end{figure}
%
In general, we observe that the
difference between global suppression and resolved suppression is not large,
i.e.\ the data points agree with the resolved suppression almost as well as 
with the global suppression. 

In Figs.\ 8a and b the difference between `H1 2006 fit B' and `H1 2006 fit A'
%
\begin{figure}
 \centering
 \includegraphics[width=0.325\columnwidth]{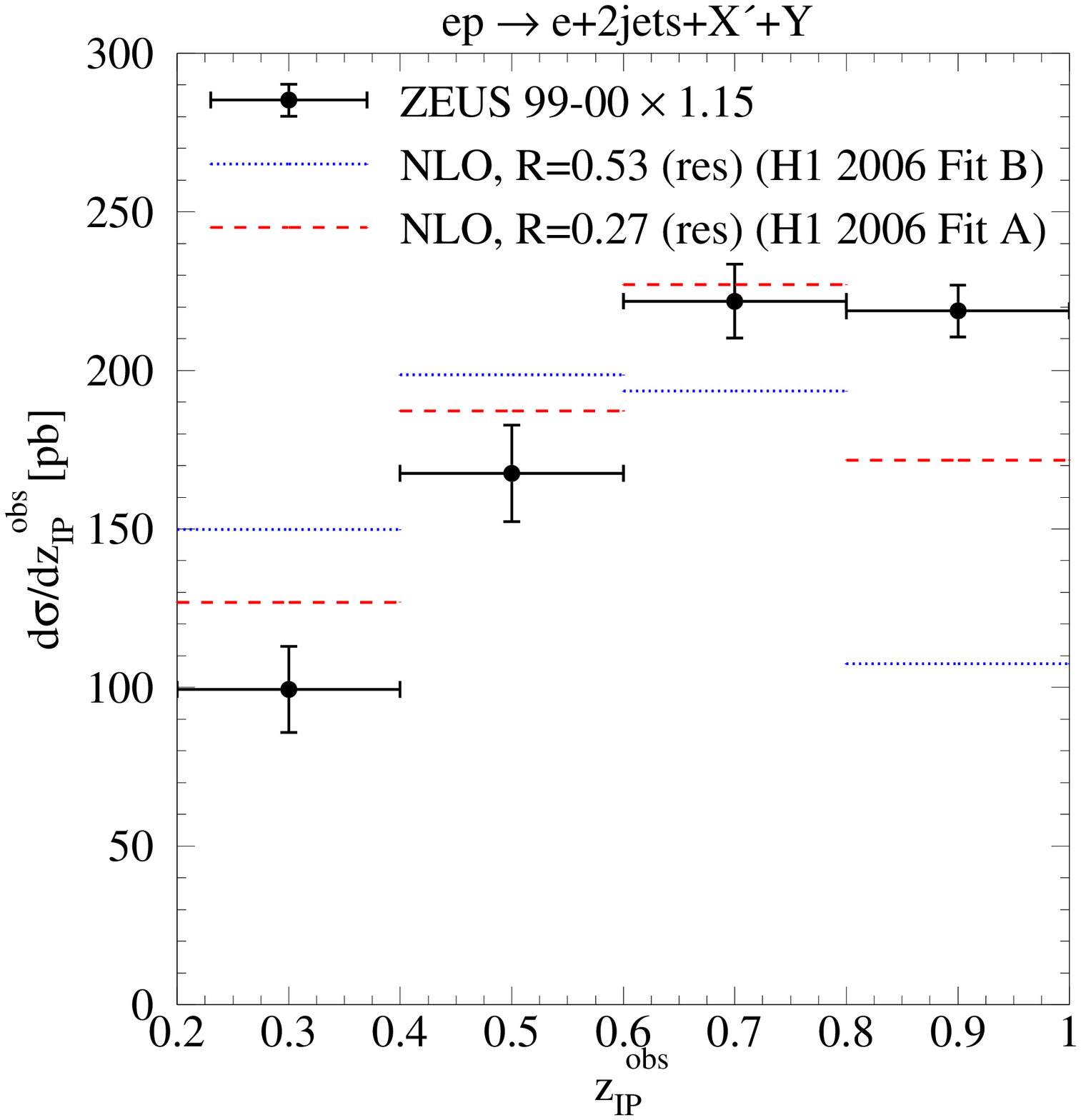}
 \includegraphics[width=0.325\columnwidth]{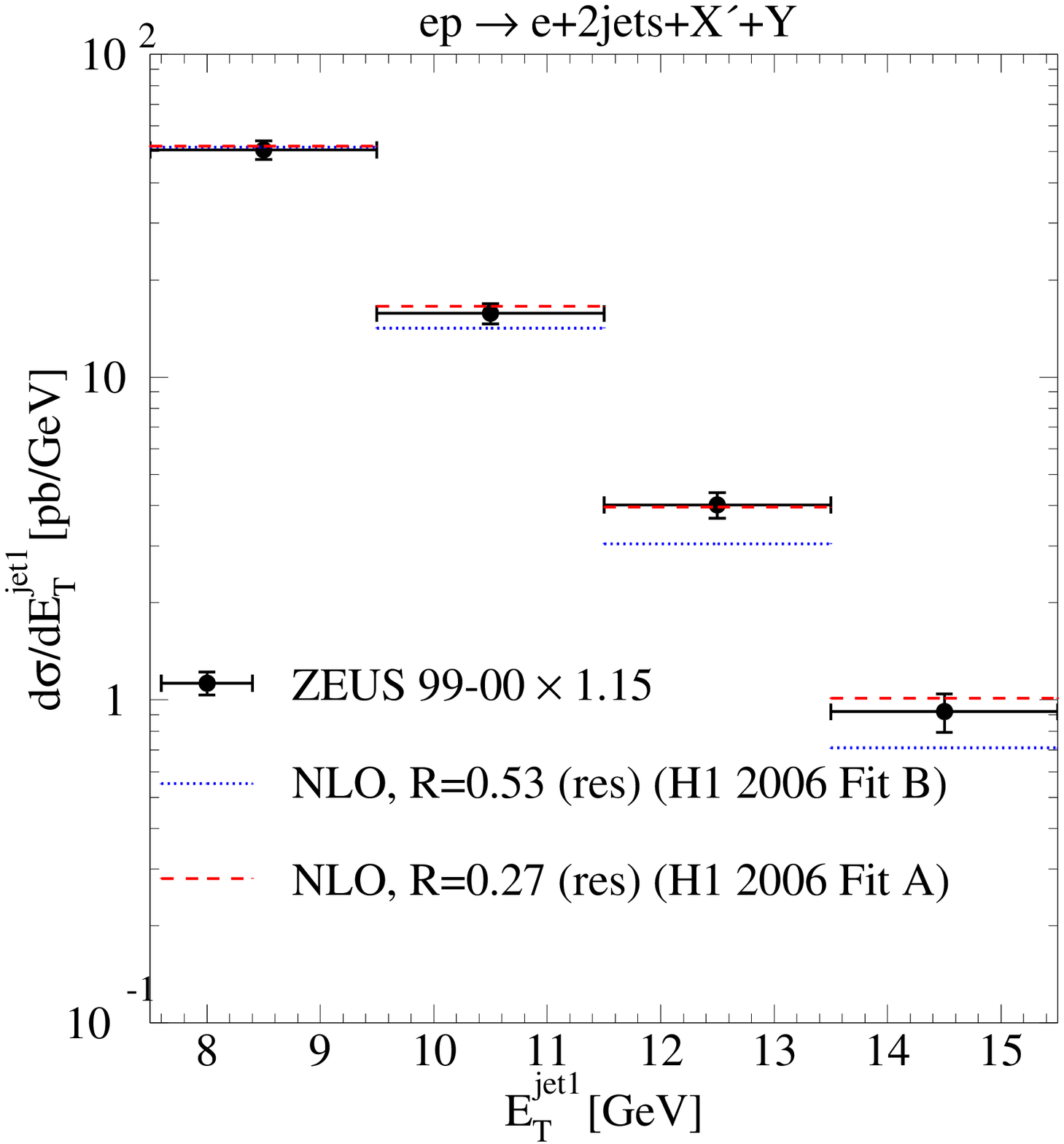}
 \caption{\label{fig:10}Differential cross sections for diffractive dijet
 photoproduction as measured by ZEUS and compared to
 NLO QCD with resolved suppression and two different DPDFs.}
\end{figure}
%
is shown again for the case of the resolved suppression. In both figures we
observe that the fit A suppression with the suppression factor $R = 0.27$
agrees better with the data than with the factor $R = 0.53$ for the
fit B suppression. In particular, for $d\sigma/dE_T^{jet1}$ the agreement with 
the three data points is perfect (note the logarithmic scale).

%
%
%

\section{Conclusion}

In summary, we have revisited the final H1 and ZEUS data on the diffractive
photoproduction of dijets at HERA. We focused on the question if the two data
sets, taken with different $ep$ center-of-mass energies and kinematic cuts (in
particular on the jet transverse energies), could be consistently interpreted
within QCD factorization, employing universal parton densities in the
diffractive exchange and process-specific hard partonic cross sections evaluated
at NLO, or showed rather evidence of factorization breaking in the direct and/or
resolved photon channels.

First, we found that even with the most optimistic (and likely realistic) parton
density set `H1 2006 fit B', both the H1 and ZEUS data sets were
overestimated by the unsuppressed NLO predictions and better described by
global suppression factors of $0.42\pm0.06$ and $0.71\pm0.06$, respectively.
These factors were obtained by fitting our NLO predictions in both cases to the
lowest (and dominant) $E_T^{jet}$-bin and are in agreement with the
global suppression factor of $0.50\pm0.10$ found by the H1 collaboration in a
fit to all of their data points, but at variance with the final conclusion in
the publication by the ZEUS collaboration.

Second, we demonstrated that the H1 (ZEUS) data (in particular the
$E_T$-distributions and somewhat less the $x_\gamma$-distributions, which are
unfortunately subject to large hadronization uncertainties) can be described
almost equally well by applying a suppression factor of about one-third (one-half)
to the resolved-only contribution. We showed that this could be consistently done
by suppressing also the direct intial-state singular part without a big impact
on the suppression factor with the added advantage of preserved
factorization-scale invariance.
Alternatively, we admitted for the
possibility that a global suppression factor might be $E_T^{jet}$-dependent,
although a theoretical motivation is only known for the first scenario and the
suppression factor obtained is in good agreement with absorptive-model
predictions.

Finally, we showed that our conclusions, while the numerical values of the
suppression factors may change to some extent, are not qualitatively altered,
when a different set of diffractive parton densities (e.g.\ `H1 2006 fit A') is
employed. The same observation should hold for the very recent 'H1 2007 fit jets',
which is very similar to the 'H1 2006 fit B'.

While the epoch of HERA experiments has now ended and an International
Linear Collider may not be built in the near future, it will be very interesting
to investigate diffractive physics at the LHC. As stated above, the search for
Higgs bosons may benefit in an important way from the diffractive production
channels, and this depends crucially on an excellent understanding of the QCD
backgrounds. Proton-proton and heavy-ion collisions at the LHC may even be
a source of high-energy photon collisions, and this may open up a whole new
field of investigation for diffractive dijet photoproduction
\cite{Klasen:2008ja}.

\end{document}